\documentclass[aps,prb,twocolumn,showpacs]{revtex4}
\usepackage[dvips]{graphicx}

\usepackage{amssymb,amsmath,amsfonts}
\usepackage{wasysym}
\usepackage{latexsym}
\usepackage{subfigure}

\newcommand{\nio}{Na$_4$Ir$_3$O$_8$}
\newcommand{\hk}{hyper-kagom\'{e}}

\begin{document}

\title{Spin-orbit effects in Na$_4$Ir$_3$O$_8$, a hyper-kagom\'{e}\, lattice
  antiferromagnet} 
\date{\today}

\author{Gang Chen}
\affiliation{Department of Physics, University of California, Santa
  Barbara, CA 93106-9530}
\author{Leon Balents}
\affiliation{Kavli Institute for Theoretical Physics, University of California,
  Santa Barbara,
  CA 93106,
  USA}

\date{\today}

\begin{abstract}

  We consider spin-orbit coupling effects in Na$_4$Ir$_3$O$_8$, a material in which
  Ir$^{4+}$ spins form an hyper-kagom\'{e}\, lattice, a three-dimensional network
  of corner-sharing triangles.  We argue that both low temperature
  thermodynamic measurements and the impurity susceptibility induced by
  dilute substitution of Ti for Ir are suggestive of significant
  spin-orbit effects.  Because of uncertainties in the crystal-field
  parameters, we consider two limits in which the spin-orbit coupling is
  either weak or strong compared to the non-cubic atomic splittings.  A
  semi-microscopic calculation of the exchange Hamiltonian confirms that
  indeed large antisymmetric Dzyaloshinskii-Moriya (DM) and/or symmetric
  exchange anisotropy may be present.  In the strong spin-orbit limit,
  the Ir-O-Ir superexchange contribution consists of {\sl unfrustrated}
  strong symmetric exchange anisotropy, and we suggest that spin-liquid
  behavior is unlikely.  In the weak spin-orbit limit, and for strong
  spin-orbit and direct Ir-Ir exchange, the Hamiltonian consists of
  Heisenberg and DM interactions.  The DM coupling is parametrized by a
  three component DM vector (which must be determined empirically).  For
  a range of orientation of this vector, frustration is relieved and an
  ordered state occurs.  For other orientations, even the classical
  ground states are very complex.  We perform spin-wave and exact
  diagonalization calculations which suggest the persistence of a
  quantum spin liquid in the latter regime.  Applications to Na$_4$Ir$_3$O$_8$ and
  broader implications are discussed.
\end{abstract}
\date{\today}
\pacs{75.10.-b,75.10.Jm,75.25.+z}




\maketitle

\section{Introduction}\label{intro}

Geometrically frustrated antiferromagnetism is a rich subject enjoying
considerable theoretical and experimental attention over several decades
of research.\cite{moessner:cjp,Ramirez:arms} Such systems are realized
by materials containing magnetic ions in which the strongest
antiferromagnetic exchanges occur on a network of bonds containing many
triangular units.  The most celebrated examples are the two-dimensional
kagom\'{e} (corner-sharing triangles) lattice and three-dimensional
pyrochlore (corner-sharing tetrahedron) lattice. In ideal classical
models, these lattices support highly degenerate ground states which
prevent order down to very low temperature.  Instead, the spins continue
to fluctuate strongly despite significant correlations induced by the
frustrated interactions.  Systems in this regime are dubbed (classical)
spin liquids, or cooperative paramagnets.  A major goal in the field is
to ascertain whether such spin liquids might also occur even in the zero
temperature limit, in which both quantum effects and many non-ideal
features of the materials must be taken into account.  The answer to
this question is quite subtle, due to many competing effects that can
come into play. Quantum and thermal fluctuations may break the ground
state degeneracy and actually induce magnetic order, an effect known as
order-by-disorder.\cite{gvozd:jetp,henley:jap,villain:jdp,bergman:natp}
This effect, however, is understood theoretically only in the large
spin, $S \gg 1$ limit, in which spins behave semi-classically.
Nevertheless, some models even with the smallest possible spins,
$S=1/2$, seem at least qualitatively to follow the order-by-disorder
scenario.  Conversely, in other models with small spin, quantum spin
liquids have been shown to occur.  No general theory to predict which of
these two tendencies obtains exists at present.  

\begin{figure}[t]
  \centering
     \includegraphics[width=2.0in]{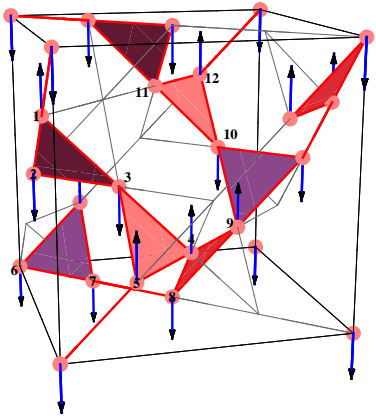}
     \caption{The \hk\ lattice of Ir$^{4+}$ spins, with one classical
       ground state of the strong anisotropy Hamiltonian shown.  Though
       this particular ground state is collinear, other ground states
       are not.}
  \label{fig:classSO}
\end{figure}

Despite this lack of theoretical discrimination, experimentalists have
forged onward in recent years, uncovering a number of promising
candidate quantum spin liquid materials with small spin $S=1/2$ on
geometrically frustrated lattices.  These include an organic magnet,
$\kappa$-(ET)$_2$Cu$_2$(CN)$_3$, containing spin-$1/2$ moments on a
slightly spatially anisotropic triangular lattice,
ZnCu$_3$(OH)$_6$Cl$_2$, an inorganic realization of a spatially
isotropic spin-$1/2$ kagome antiferromagnet, and very recently the cubic
material \nio\, which realizes an \hk\ antiferromagnet, in which
spin-$1/2$ moments reside on a three-dimensional network of
corner-sharing triangles\cite{takagi:prl} -- see Fig.~\ref{fig:classSO}.
None of these compounds exhibit indications of magnetic ordering.  The
interpretation of the first two materials, however, is complicated by
the appearance of inhomogeneous magnetic moments at low temperature in
$\kappa$-(ET)$_2$Cu$_2$(CN)$_3$ \cite{shimizu:prb}, and by fairly high
levels of substitutional disorder (Zn for Cu) in ZnCu$_3$(OH)$_6$Cl$_2$.
By contrast, the Ir$^{4+}$ moments are expected to be well ordered in
\nio\ , due to the much larger ionic radius of Ir compared to Na and O.

Recent two works\cite{hopkinson:prl,lawler:07} assumed the nearest 
neighbor antiferromagnetic Heisenberg model
for \nio. In Ref.~\onlinecite{hopkinson:prl}, 
the authors treated the spin as a classical $O(N)$ spin. By a large-$N$ mean field 
theory and classical Monte Carlo simulation, they found that the classical 
ground states are highly degenerate and a nematic order emerges at low 
temperatures in the Heisenberg model ($N=3$) via ``order by disorder'', 
representing the dominance of coplanar spin configuration. 
In Ref.~\onlinecite{lawler:07}, the authors presented a large-$N$ $Sp(N)$ 
method and studied both the semi-classical spin and quantum spin regimes. 
In the semi-classical limit, they predicted that an unusual $\vec{k}=(0,0,0)$ 
coplanar magnetically ordered ground state is stabilized with no local 
``weather vane'' modes. While in the quantum limit, a gapped topological 
$Z_2$ spin liquid emerges.  

Due to the large atomic number ($Z=77$) of Ir, however, we should
carefully consider the role of spin-orbit coupling, whose leading effect
in localized $S=1/2$ electron systems is the Dzyaloshinskii-Moriya (DM)
interaction in the weak spin-orbit coupling limt.  
In fact, DM interactions have been argued to play an
important role even in ZnCu$_3$(OH)$_6$Cl$_2$, with much less
relativistic Cu ($Z=29$) moments.\cite{rigol:prb} The DM interaction
reduces the full SU(2) spin-rotational invariance of the Heisenberg
Hamiltonian to the $Z_2$ discrete time-reversal symmetry (in addition to
coupling spin transformations to the discrete point group operations of
the lattice).  On general grounds, this is expected to lower the
degeneracy of the classical ground state manifold.  However, depending
upon the detailed form of the DM coupling, varying degrees of degeneracy
remain, indicative of different amounts of frustration.  The tendency of
the system to retain the order of the classical ground state is
certainly also variable, and warrants investigation.  This is one of the
motivations of the present study.

Another motivation comes directly from the experiments in
Ref.\onlinecite{takagi:prl}, several aspects of which are suggestive of
the presence of spin-orbit coupling.  First, the ``Wilson ratio'' $R = T
\chi/c_v$, is observed to {\sl grow} with cooling at low temperature,
following a power-law $R \sim 1/T^{\alpha-1}$, with $2<\alpha<3$.  Here
$\chi \sim {\rm const.}$ is the magnetic susceptibility and $c_v \sim
T^\alpha$ is the specific heat.  As will be discussed in
Sec.~\ref{sec:therm-spin-rotat}, such a low temperature behavior is
incompatible with {\sl any} spin-rotationally invariant phase of matter
supporting well-defined quasi-particle excitations.  To our knowledge,
it is at odds with all known theoretical models of quantum spin liquids,
and seems highly unlikely on general grounds.  Taking into account the
observed field-independence (up to 12 Tesla) of the specific heat $c_v$
brings the behavior even further into disagreement with
spin-rotationally invariant theories.  Second, samples in which a
fraction $x$ of Ir atoms are substituted by Ti (which are in a
non-magnetic Ti$^{4+}$ state) display a Curie component in the
susceptibility linearly proportional to $x$ with a {\sl strongly
  suppressed amplitude}, of approximately $1/3$ of a spin-$1/2$ moment
per Ti.  As we also show in Sec.~\ref{sec:impur-susc}, such behavior is
also at odds with any {\sl simple} spin-rotationally invariant
low-temperature phase (assuming no clustering of the Ti atoms), though
some more exotic.  All these observations, however, are readily
reconciled by assuming the presence of spin-rotational symmetry
breaking.  Given the lack of any observed magnetic ordering, explicit
and substantial spin-orbit interactions would appear to be a likely candidate.

In Sec.~\ref{sec:symm-micro}, we consider an explicit semi-microscopic
calculation of the exchange Hamiltonian in the presence of spin-orbit
coupling.  We consider both super-exchange through the intermediate O
atoms, and direct exchange between closest pairs of Ir spins.  The
results depend crucially upon the relative magnitude of the spin orbit
coupling constant $\lambda$ and the non-cubic splittings of the $t_{2g}$
multiplet.  This is quantified by two dimensionless ratios of $\lambda$
to the two energy splittings $\epsilon_{2}-\epsilon_1$ and
$\epsilon_3-\epsilon_1$ of the orbital levels in the absence of
spin-orbit.  When $\lambda$ is the largest energy scale -- the ``strong
spin orbit limit'' -- the ``spin'' has a substantial orbital angular
momentum component, while in the opposite ``weak spin orbit limit'', it
is predominantly microscopic spin angular momentum.  Indeed, the
$g$-factor has {\sl opposite sign} in the two limits.  Which if either
limit applies is the most fundamental physical question to be understood
concerning the nature of magnetism in \nio.   We are not aware of any
calculations or direct experimental measurements that indicate whether
\nio\ is in the weak or strong spin-orbit limits, or intermediate
between these situations.  Instead we will address this question by
comparing the expected phenomenology for the two cases to experimental
observations.

In the strong spin-orbit limit, when the dominant mechanism is Ir-O-Ir
superexchange, we find an highly anisotropic effective spin Hamiltonian,
in which two spin components on each bond interact antiferromagnetically
while the third interacts ferromagnetically.  Specifically,
\begin{equation}
  \label{eq:1}
  {\mathcal H} = \sum_{\langle i j \rangle} J \epsilon^\mu_{ij} S_i^\mu S_j^\mu,
\end{equation}
where $(\epsilon_{ij}^x,\epsilon_{ij}^y,\epsilon_{ij}^z)$ is a
permutation of $(+1,+1,-1)$ chosen appropriately for each bond (see
Sec.~\ref{sec:strong}) to specify the two antiferromagnetic and one
ferromagnetic direction. We call Eq.~(\ref{eq:1}) the ``strong
anisotropy'' Hamiltonian.  

Somewhat surprisingly, the remaining three cases: strong spin orbit and
direct exchange, and weak spin-orbit and superexchange {\sl or} direct
exchange, all lead to approximately isotropic Heisenberg interactions.
For the weak spin-orbit limit, this is guaranteed, but it is certainly
not in the strong spin-orbit case.  The dominant spin-rotational
symmetry breaking effect, which is perturbative in all three regimes, is
the DM interaction.  The effective Hamiltonian has the form
\begin{equation}
  {\mathcal H} = \sum_{\langle i j \rangle} [ J {\bf S}_i \cdot {\bf S}_j +
{\bf D}_{ij} \cdot ({\bf S}_i \times {\bf S}_j) ]
\; .
\label{eq:hamiltonian}
\end{equation}
Here $J$ is the same for all bonds, and estimated as $J\approx {\rm
  400}K$ from the measured Curie-Weiss temperature $\Theta_{CW} \approx
-{\rm 650}K$.  Symmetry strongly restricts the structure of this
effective magnetic Hamiltonian for \hk.  The full set of DM vectors
${\bf D}_{ij}$ may be fixed by just three parameters.  That is, ${\bf
  D}_{ij}$ on any one bond is arbitrary (by symmetry), but that choice
determines all remaining ${\bf D}_{ij}$ in the system.  It is convenient
to choose the local coordinate system $(D_1, D_2, D_3)$, where $D_1$ is
the component aligned with the bond, $D_2$ is normal to the triangle
plane in \hk\, lattice, $D_3$ lies in the triangle plane but
perpendicular to the bond (see Fig.~\ref{fig:dm}).  The semi-microscopic
calculations in Sec.~\ref{sec:estimates} confirm that all three
components are non-vanishing, and gives a quantitative understanding of
them.  Due to the large $\lambda$ and considerable uncertainties in
estimating the non-cubic energy splittings, it is difficult to estimate
the overall magnitude of the DM terms, but there is no reason they need
be particularly small, though the perturbative estimates are presumably
valid only for $|D_i|<J \lesssim 0.1$ or so.  A na\"ive estimation is
obtained by noting that in this limit the ratios of DM to exchange are
expected to be of the same order as the shift of the $g$-factor,
i.e. $|D_i|/J \sim |1-|g|/2|$.  From the measured moment $\mu_{eff} \approx
1.9\mu_B = |g|\mu_B$, assuming we are in this limit would give $|D_i|/J
\sim 0.05$ or so.
In Sec.~\ref{sec:strongspin}, we considered the strong anisotropy
Hamiltonian, Eq.~(\ref{eq:1}) in the classical approximation.
Remarkably, unlike the Heisenberg model which is macroscopically
degenerate (i.e. its ground states are specified by a number of
continuous parameters proportional to the number of spins), the system
in this limit has an almost unique ground state.  We find a continuous
{\sl two parameter} manifold of ground states, in which any one spin can
be specified arbitrarily after which all others are determined.  This is
still a (small) accidental degeneracy, since the system has itself only
discrete (space-group and time-reversal) symmetries which do not protect
any continuous degeneracies.  Nevertheless, this degeneracy is
presumably insufficient to prevent ordering in a classical system.  The
behavior in the physical $S=1/2$ quantum problem is not known, but one
would expect that an ordered phase of the same symmetry as the classical
one is rather likely, and there is little reason to suppose a
significant suppression of the ordering temperature relative to the
Curie-Weiss scale.  The disagreement of these expectations with the
experimental observations suggests that it is the weakly-anisotropic DM
Hamiltonian rather than this one which is most appropriate.  We however
return to this question in more detail in Sec.~\ref{sec:discussion}.
\begin{figure}
	\centering
		\includegraphics[width=2.3in]{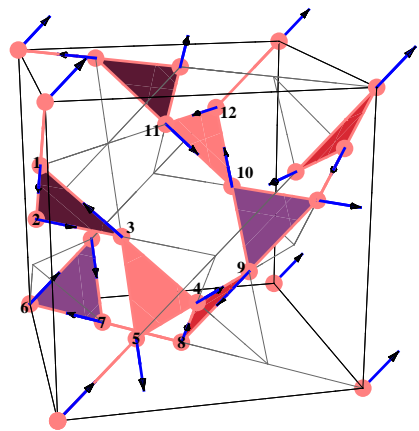}
                \caption{(Color online) The ``windmill'' state, which is
                  the classical ground state in the weak anisotropy
                  limit when $D_2<0$.  It is also the basis vector
                  $\psi_2 $ (Table.~\ref{tab:basis}) of one dimensional
                  representation $\Gamma_2^{(1)}$ (see
                  Eq.~\eqref{eq:dec}).  In the generic system with
                  non-zero $D_1,D_3$, the spins are slightly canted out
                  of the plane of each triangle.}
	\label{fig:hkorder1}
\end{figure}

In Sec.~\ref{sec:order}, we turn to the weak anisotropy limit, and
first explore the classical phase diagram of Eq.~(\ref{eq:hamiltonian}).
In general, even this optimization problem is highly non-trivial, given
the large unit cell of the \hk\ lattice, and the possibility that the
magnetic unit cell of the ground states may be yet larger.  In the
special case $D_1=D_3=0$ and $D_2<0$, however, it is possible to solve
this problem exactly.  The degeneracy is broken {\sl completely} to a
single Kramer's pair of coplanar ground states, for which the magnetic
unit cell is equal to the crystallographic one.  These may in this sense
be considered ${\bf k}=(0,0,0)$ states.  One is drawn in
Fig.\ref{fig:hkorder1}.  We call this the ``windmill'' state.  By several
approximate methods, we establish the form of the phase diagram in the
general $D_1$-$D_2$-$D_3$ parameter space.  Generically the windmill
state distorts to a ``canted windmill'' state (still with ${\bf
  k}=(0,0,0)$), occupying a finite region of the phase diagram.  In
addition, one finds a wide range of incommensurate phase, in which the
ordering wavevector ${\bf k}$ is non-zero and generically irrational in
reciprocal lattice coordinates.  Owing to the breaking of space group
symmetries, the incommensurate phase retains more of the
frustration-induced degeneracy.

A key question is whether the DM interactions, expected on physical
grounds and invoked phenomenologically to explain the experimental
properties discussed above, are consistent with the observed spin liquid
behavior of \nio, i.e. the lack of any ordering down to the very low
temperatures of $T \approx 1.8K = \Theta_{CW}/360$.  The breaking of
degeneracy by DM might be expected to reduce quantum fluctuations and
thereby lead to ordering, in conflict with experiment.  To study this
possibility, we carried out spin wave calculations of the excitation gap
and the quantum correction to the classical ordered moment.  Indeed, we
find that deep inside the ${\bf k}=(0,0,0)$ phases, the quantum
correction is not too large, which leads us to expect that the
spin-$1/2$ system exhibits the classical order.  However, we find very
large quantum corrections elsewhere in the phase diagram, even for
fairly substantial $|D_i|$. In our results, small excitation gap will
lead to a large quantum correction to classical ordered moment.
Decreasing the excitation gap by changing the DM vector will eventually
destroy the classical ordered moment completely. In this regime, the
large quantum effects invalidate the spin-wave treatment and indeed
leave open the possibility of a quantum spin liquid, consistent with
experiment.  To further confirm the results and treatment of spin wave
theory, we implemented exact diagonalization on a small cluster (six
triangles with $13$ spins). The excitation gap obtained from numerical
data of specific heat qualitatively agrees with the prediction of spin
wave theory.

The remainder of this paper is organized as follows. In
Sec.~\ref{sec:symm-micro} we discuss the symmetry allowed DM vector
components and calculate the exchange spin Hamiltonian with a
microscopic theory for both strong and weak spin-orbit coupling.  In
Sec.~\ref{sec:strongspin} we discuss the classical ground states of the
strong anisotropic exchange Hamiltonian obtained from Ir-O-Ir
superexchange in the strong spin-orbit coupling limit.  In
Sec.~\ref{sec:order} we will turn to look at the weak anisotropy
Hamiltonian, namely, the nearest-neighbor Heisenberg model with small DM
interactions.  We first present the magnetic ordered state when $D_2 <
0$ then discuss the more general case when nonvanishing $D_1$ and $D_3$
components are present in the system. In Sec.~\ref{sec:spinwave}, we
present a linear spin wave theory to find the zero temperature quantum
correction to the magnetically ordered phase and compare with exact
diagonalization. Finally, a discussion of our main results and their
relevance to \nio\, is given in Sec.~\ref{sec:discussion}.

\section{Thermodynamics of spin-rotationally invariant magnetic
     phases}
\label{sec:therm-spin-rotat}

In this section, we discuss some apparent constraints on the low
temperature susceptibility and specific heat in spin-rotationally
invariant phases of matter.  As described in the introduction, these
constraints appear to be violated in \nio, which we take as an
indication of the presence of substantial spin-orbit interactions.

\subsection{Clean system}
\label{sec:clean-system}

We take spin-rotational invariance to mean the existence of global SU(2)
spin symmetry.  According to standard quantum mechanics, this implies
that all states may be chosen as eigenstates of $S_{\rm TOT}^2$ and
$S_{\rm TOT}^z$, where $\vec{S}_{\rm TOT}$ is the operator for total
spin.  The choice of $z$ axis being arbitrary, we take it along the axis
of any applied field.  The effect of the field on the system is then
entirely described by the term
\begin{equation}
  \label{eq:2}
  {\mathcal H}_H = - H \sum_i S_i^z = - H S_{\rm TOT}^z
  \;,
\end{equation}
where we have absorbed the (presumed known) $g$-factor, Bohr magneton,
etc. into the definition of $H$.  One observes from Eq.~(\ref{eq:2})
that ${\mathcal H}_H$ is diagonal in the $S^z_{\rm TOT}$ basis, and thus
the Hamiltonian eigenstates themselves {\sl are independent of field},
and only the eigenvalues change.  Focusing on the states rather than
their energies, we may say that the only effect of the field upon the
system in equilibrium is to modify the occupation probabilities of
states.  In this sense, the magnetic field is a thermodynamic
perturbation, and the susceptibility is a thermodynamic quantity,
determined only by the density of states.  The specific heat is of
course also such a thermodynamic quantity, determined from the same
density of states.  Thus they are connected.

Specifically, the specific heat is a measure of the full density of
states for all excitations above the ground state, irrespective of their
spin quantum numbers.  The susceptibility, however, only counts those
excitations which carry non-zero spin $S^z$ along the field.  The
possibility of spin-less excitations allows some independence of the
two: by introducing more $S^z=0$ states, one can increase $c_v$
arbitrarily while leaving $\chi$ unchanged.  However, the converse is
not true.  It would seem difficult to increase $\chi$ without also
contributing to $c_v$.  The only way in which this can be done is to
introduce states with very large $S^z$ (which then contribute a large
amount to $\chi$) but very low energy (and hence do not contribute much
to $c_v$).  This case corresponds to a system on the verge of a
ferromagnetic instability.

Without fine-tuning to such a point, we are led to expect that, in the
presence of SU(2) symmetry,  the
Wilson ratio,
\begin{equation}
  \label{eq:3}
  R = \frac{T\chi}{c_v}
\end{equation}
should have an {\sl upper} bound, corresponding to all excitations
contributing both to $\chi$ and $c_v$.  This can indeed be shown
provided we assume the system can be described by a non-magnetic ground
state and non-interacting quasiparticles characterized by a spin $S^z$
quantum number.  We define the density of state $g^b_m(\epsilon)$ and
$g^f_m(\epsilon)$ for boson or fermion excitations carrying spin
$S^z=m$, respectively.  The specific heat is
\begin{equation}
  \label{eq:4}
  c_v = \partial_T \sum_m \int_0^\infty \!\!\! d\epsilon \epsilon
  \left[g_m^b(\epsilon) n_b(\epsilon) + g_m^f(\epsilon)n_f(\epsilon)\right]
  \;,
\end{equation}
where
\begin{equation}
  \label{eq:5}
 n_{b/f}(\epsilon) = \frac{1}{e^{\beta \epsilon}\mp 1}
 \;.
\end{equation}
One obtains
\begin{equation}
  \label{eq:6}
  c_v = \frac{k_B^2 T}{4} \sum_m \int_0^\infty \!\!\! dx x^2 \left[
    \frac{g_m^b(k_B T x)}{\sinh^2 (x/2)} + \frac{g_m^f(k_B T x)}{\cosh^2
      (x/2)}\right]
      \;.
\end{equation}
Now consider the susceptibility
\begin{eqnarray}
  \label{eq:7}
  \chi & = & \partial_H \sum_m \int_0^\infty \!\!\!
    d\epsilon m
  \big[g_m^b(\epsilon) n_b(\epsilon-H m) \\
& &  \hspace{0.6in} +
    g_m^f(\epsilon)n_f(\epsilon-H m)\big]\Big|_{H=0}
     \nonumber
     \;.
\end{eqnarray}
One finds
\begin{equation}
  \label{eq:8}
  \chi =  \frac{1}{4} \sum_m m^2 \int_0^\infty \!\!\! dx \left[
    \frac{g_m^b(k_B T x)}{\sinh^2 (x/2)} + \frac{g_m^f(k_B T x)}{\cosh^2
      (x/2)}\right]
      \;.
\end{equation}
In the low temperature limit, we may approximate $g_m^{f/b}(k_B T x)$ by
its small argument behavior, which is usually a power-law form:
\begin{equation}
  \label{eq:9}
  g_m^{b/f}(\epsilon) \sim A_m^{b/f} \epsilon^{\gamma_m^{b/f}}
  \;.
\end{equation}
One needs obviously $\gamma_m^{f/b}>-1$ for the density of states to be
integrable (and hence the cumulative distribution well defined).  We
will encounter problems with Eq.~(\ref{eq:8}) if $\gamma_m^b \leq 1$ for
any $m\neq 0$.  This could be fixed by the inclusion of a chemical
potential, whose temperature dependence we have ignored, and as usual is
necessary to avoid Bose condensation of free bosons at low $T$ when
their energy is close to zero.  This effect, however, does not change
any of the results, so we have excluded it for simplicity here.  

Given Eq.~(\ref{eq:9}), the specific heat will be controlled at low $T$
by the minimum exponent over {\sl all} $\gamma_m^{b/f}$:
\begin{equation}
  \label{eq:11}
  \gamma_0 = {\rm min} \left\{ \gamma_m^{b/f} \right\}
  \;.
\end{equation}
One has
\begin{equation}
  \label{eq:10}
  c_v \sim A_0 k_B^{2+\gamma_0} T^{1+\gamma_0}
  \;,
\end{equation}
with some constant $A_0$.  The susceptibility is controlled by the
minimum exponent for $m\neq 0$:
\begin{equation}
  \label{eq:12}
   \gamma_1 = {\rm min} \left\{ \gamma_m^{b/f} ; \qquad m \neq 0 \right\}
   \;.
\end{equation}
Note that by definition, $\gamma_0 \leq \gamma_1$.
Then
\begin{equation}
  \label{eq:13}
  \chi \sim A_1 T^{\gamma_1}
  \;.
\end{equation}
Then the Wilson ratio becomes
\begin{equation}
  \label{eq:14}
  R \sim R_0 T^\Upsilon
  \;,
\end{equation}
where $R_0 = \frac{A_1}{A_0  k_B^{2+\gamma_0}} $ and
\begin{equation}
  \label{eq:15}
  \Upsilon = \gamma_1-\gamma_0 \geq 0
  \;.
\end{equation}

Because $\Upsilon \geq 0$, the Wilson ratio cannot diverge on lowering
$T$, and unless $\Upsilon=0$, actually vanishes as $T\rightarrow 0$.  In
defining the Wilson ratio, we have considered only the zero field
specific heat.  In a field, contributions from all excitations with
$m\neq 0$ will be field dependent.  So unless the $m=0$ mode is dominant
in $c_v$, the specific heat should be expected to be field dependent.
Conversely, field independence of the specific heat requires that the
$m=0$ excitations dominate $c_v$.  In this case, we have $\Upsilon >0$,
and the equality is {\sl not} satisfied.  Thus a field-independent
low-temperature specific heat would be expected to correspond to a {\sl
  vanishing} Wilson ratio as $T\rightarrow 0$.  This makes the observed
{\sl divergence} of $R$ on lowering $T$ in \nio\ even more at odds with
the theoretical expectation for an SU(2) invariant system.

A few comments are in order.  First, while we have assumed power-law
forms for the low energy density of states, this is not essential.  We
believe the lack of low temperature divergence in $R(T)$ is very robust
within the quasiparticle picture.  Beyond the quasiparticle
approximation, the situation is less clear, and we do not have a
definitive proof of this behavior of $R(T)$.  However, we do not know of
any single theoretical counter-examples in the literature for SU(2)
invariant low temperature phases.  

If SU(2) symmetry (or more specifically, invariance under spin rotations
about the measurement axis) is broken, however, one readily and indeed
almost generically observes this behavior.  This is quite familiar from
the case of ordered antiferromagnets in two or three dimensions.  These
are well-known to display a non-vanishing constant zero temperature
uniform susceptibility $\chi_0$ and a power-law specific heat $c_v \sim
A T^d$ due to spin wave excitations, hence a Wilson ratio obeying
Eq.~(\ref{eq:14}) with however $\Upsilon = 1-d <0$.  This arises because
the ground state itself is modified continuously by the introduction of
a magnetic field.  Semi-classically, the magnetic field leads to a
smooth canting of the antiferromagnetic moments in the field direction,
linearly proportional to the applied field.

This phenomena is, however, not limited to systems with spontaneous
symmetry breaking.  It occurs whenever the effective Hamiltonian for the
low temperature phase does not conserve the spin component along the
magnetic field.  As an extreme example, one may consider the case of two
spin-$1/2$ spins coupled together by antiferromagnetic exchange and DM
interaction:
\begin{equation}
  \label{eq:16}
  {\mathcal H}_2 = J {\bf S}_1\cdot {\bf S}_2 - D {\bf\hat z}\cdot {\bf
    S}_1 \times {\bf S}_2 - H \left( S_1^x + S_2^x\right)
    \;,
\end{equation}
where we have chosen the DM vector along the $z$ axis, and therefore
oriented the field along $x$ so that it couples to a non-conserved
magnetization.  One can readily diagonalize the Hamiltonian, and find
that in zero field has a unique ground state with a gap $\Delta=1/2(J +
\sqrt{J^2+D^2})$.  Nevertheless, the susceptibility is non-zero when
$D\neq 0$:
\begin{equation}
  \label{eq:17}
  \chi = \left.\frac{\partial S_i^x}{\partial H}\right|_{H=0} =
  \frac{\sqrt{J^2+D^2} - J}{J\left(\sqrt{J^2+D^2}+J\right)} \;.
\end{equation}
Because of the gap, the specific heat of the dimer is activated at low
temperature, and hence the dimer's Wilson ratio diverges {\sl
  exponentially} at low temperature.  In general, a non-zero limit for
the low temperature susceptibility is always to be expected once SU(2)
symmetry breaking perturbations are taken into account.  The specific
heat, however, is insensitive to symmetry, and remains a true probe of
low energy modes.  

\subsection{Impurity susceptibility}
\label{sec:impur-susc}

In \nio, the introduction of non-magnetic impurities (substitution of
Ti$^{4+}$ for Ir$^{4+}$) was observed to give rise to a Curie component
with a reduced effective moment of $\mu_{eff} \approx (2\mu_B)/3$ per
Ti.  We would like to argue that a spin liquid state with such a large
reduction from the moment of a free spin, $2\mu_B$, is unlikely in the
absence of spin-orbit interactions, but quite likely when they are
invoked.

Suppose the Hamiltonian has global SU(2) spin-rotational symmetry in the
absence of an applied magnetic field.  Then a spin-liquid ground state,
which, by definition, does not break SU(2) symmetry, must be a spin
singlet, i.e. a state of total spin $S=0$.  Its excitations can
therefore by characterized  by spin quantum numbers.   Representations of
SU(2) always have integer or half-integer spin, and in particular for
all these the projection of the total spin along any field axis is a
multiple of $1/2$.  

Now consider a single impurity.  It may be a strong perturbation
locally, but does not perturb the Hamiltonian far from itself.  Again
presuming spin-orbit can be neglected, the ground state of this system
should be a spin eigenstate, though not necessarily non-zero.
Nevertheless, it can be classified by a total spin which is a multiple
of an half integer.  It is natural to expect that the ground state
multiplet of a single impurity controls the impurity susceptibility (but
see below).  Allowing now for an external field, this is simply
described as in the previous subsection by Eq.~(\ref{eq:2}).  Since the
low energy states are still good representations of SU(2), and only the
total spin projection enters Eq.~(\ref{eq:2}), we will obtain an
effective moment which is at a minimum (if it is non-zero) $2\mu_B$ per
impurity.

The caveat in this argument is the possibility of a Kondo-like effect.
If the spin liquid state is gapless, then there is a possibility for an
impurity moment to be ``screened'' by the bulk degrees of freedom.
Still, the possibility of a fractional impurity moment is delicate.
Most Kondo effects either completely screen the moment (as in the single
channel case, leading to $\mu_{eff}=0$) or to weaker temperature
dependence of the impurity susceptibility (e.g. $\chi_{imp}\sim |\ln
T|$ in the two-channel model, which has a non-trivial Kondo fixed
point). Thus most types of Kondo effect do not allow for such behavior.
Recently, it has been suggested that some spin liquids might sustain a
{\sl critical fixed line} of Kondo fixed points, connected to the free
impurity fixed point.  This situation can in fact lead to a renormalized
Curie constant.\cite{chen:prb} It would indeed be
appealing should such an exotic possibility be realized in \nio, but we
should allow for simpler explanations.

As is well-known, the effective moment of ions in solids varies widely
from the quantized values expected from SU(2) symmetric considerations.
This is of course due to spin-orbit coupling.  In general, with
spin-orbit interactions present, the ground state of an impurity can be
expected to be a Kramer's singlet or a Kramer's doublet.  In the latter
case, it will behave energetically (i.e. in specific heat) as a
spin-$1/2$ spin, but will have in general a non-trivial g-tensor
describing its coupling to a field.  This reflects a change in the
effective moment.  Thus there is no ``quantization'' of the effective
moment once spin-orbit coupling is substantial.  The observed fractional
effective moment in \nio\ is perhaps another indication in this direction.

\section{Spin-Orbit coupling in the \hk\ lattice}
\label{sec:symm-micro}

In this section, we discuss the form of the spin-orbit modifications to
the isotropic Heisenberg Hamiltonian.  This is not directly calculable
from semi-microscopic considerations without some assumptions about the
local energetics due to crystal field splittings.  Therefore we consider
below a number of cases.

\subsection{Symmetry allowed DM vector components}
\label{sec:symmetry-allowed-dm}

\begin{figure}[hbt]
	\subfigure{
	\label{fig:dmipath}
		\includegraphics[width=2.0in]{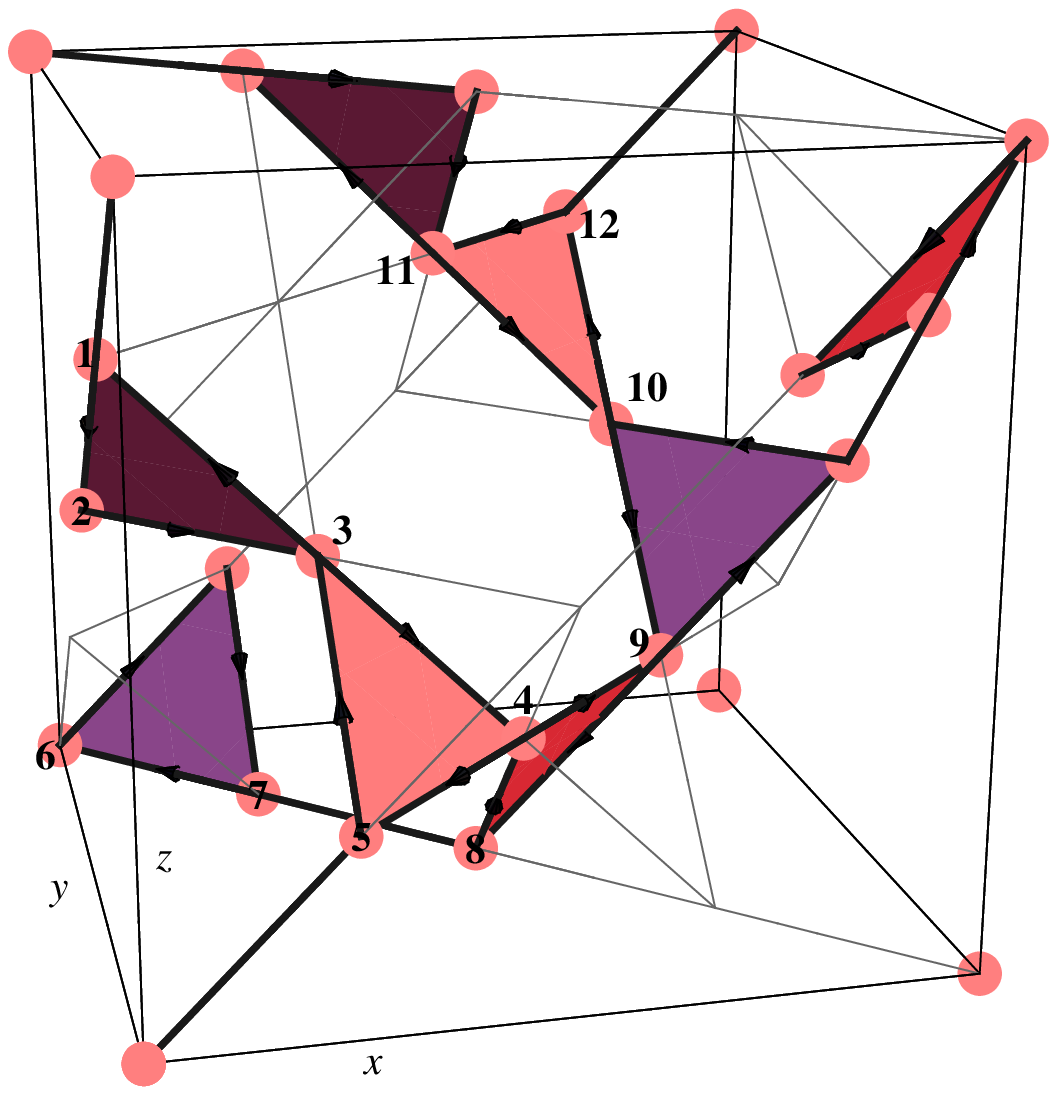}}
	\subfigure{
	\label{fig:dmdirec}
		\includegraphics[width=1.2in]{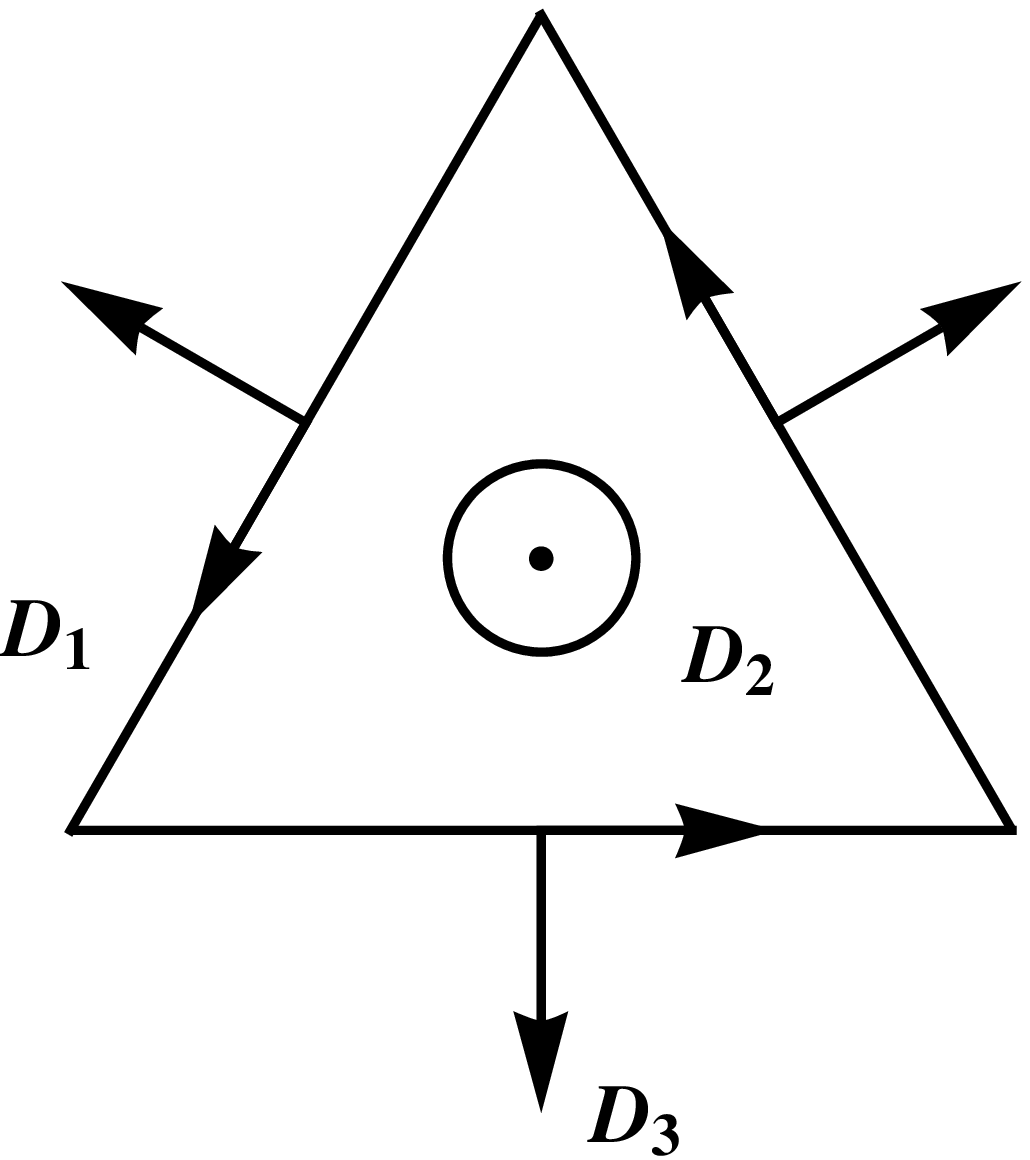}}
              \caption{(Color online) Left: One unit cell of the \hk\
                lattice.  The pink balls are occupied by magnetic ions,
                which are connected by dark black bonds. There are 12
                sites in one primitive unit cell.  The arrow from site
                $i$ to site $j$ corresponds to ${\bf D}_{ij} \cdot ({\bf
                  S}_i \times {\bf S}_j) $ in the Hamiltonian.  We will
                call these arrows DM interaction path.  Right: DM vector
                components illustrated on one triangle. $D_1$ is the
                component which is aligned with the DM interaction path
                (Left). $D_2$ is the component normal to triange
                plane. The direction is decided by the chirality of bond
                direction. $D_3$ is the component perpendicular to the
                bond but in the triangle plane.}
	\label{fig:dm}
\end{figure}

In several cases, we will find that the dominant effect of spin-orbit
coupling is to induce Dzyaloshinskii-Moriya (DM) interactions between
the nearest-neighbor spins.  Therefore before attempting any
calculations, it is instructive to first consider the symmetry
constraints upon them.  Generally, DM interactions are rather highly
constrained.  For instance, they are absent if there is an inversion
center between the two spins in question (this is not the case in \nio).
The compound \nio\ has cubic symmetry, described by the space group
P4$_1$32, and consequently has a number of point group symmetries.  For
our purposes, it is useful to consider a unconventional set of
generators of these symmetries.  Specifically, the full point group can
be generated from the set of $180^{\circ}$ rotations around a local
$C_2$ axis at each site.  Due to this symmetry, all the \hk\ sites and
bonds are equivalent.  In Table.~\ref{tab:basis}, we list the directions
of the $C_2$ axes ($\psi_1$) for every site in the unit cell (see
Fig.~\ref{fig:dm} for the labeling).  The $C_2$ rotational symmetries
relate the DM vectors of any two bonds.  That is, given the DM vector on
any one \hk\ bond, all others are determined.  This one DM vector,
however, is itself entirely unconstrained by the P4$_1$32 symmetry.

Since any single bond of the \hk\ is uniquely associated with one
triangle, it is natural to adopt a local coordinate system based on this
triangle to describe the DM vector's components.  We denote the
component aligned with the bond $D_1$, the component normal to the
triangle plane $D_2$ and the component normal to the bond but localized
in the triangle plane $D_3$. Three components have been illustrated in
Fig.~\ref{fig:dm}. If we select the direction of $D_1$ component
axis by assigning a direction to one bond (arrows in Fig.~\ref{fig:dm}),
the $C_2$ rotation symmetry can generate the equivalent $D_1$ axis for
other bonds (see Fig.~\ref{fig:dm}). In every triangle, there is a
chirality of the $D_1$ axis of three edges, which can be considered as
the direction of $D_2$ axis. The cross product of $D_1$ and $D_2$
directional vector generates the direction of $D_3$ axis.

Such a parametrization may be applied not only for the \hk\, lattice, but
for any lattice consisting of corner-sharing triangles, such as the
slightly distorted kagom\'{e} lattice of
Fe/Cr-jarosites.\cite{elhajal:prb,ballou:pss,elhajal:pb} In that
example, the $D_1$ component is forbidden by a mirror plane symmetry.
In \nio, there are as we said no constraints on the $D_i$, and we might
naively expect all three components to be non-vanishing and
comparable.  We will investigate this by microscopic calculations
below.

\subsection{Local electron energetics of Ir ion}
\label{sec:locenergy}

\begin{figure}
  \centering
		\includegraphics[width=2.5in]{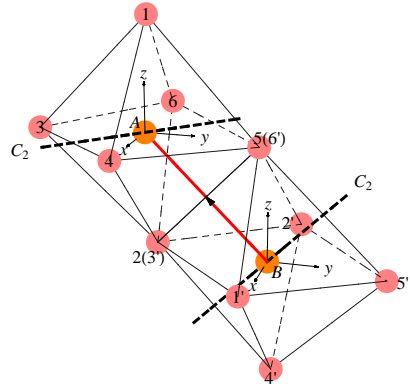}
\caption{(Color online) Ir$^{4+}$ and octahedron O$^{2-}$ environment 
(Thin black line).  Two neighboring Ir$^{4+}$ are denoted by A and B 
(In orange). A/B's six O$^{2-}$ are labelled as $1$/$1^{\prime}$, 
$2$/$2^{\prime}$, $3$/$3^{\prime}$, $4$/$4^{\prime}$, $5$/$5^{\prime}$ 
 and $6$/$6^{\prime}$ (In pink), in which, $2$ and $3^{\prime}$, 
 $5$ and $6^{\prime}$ label the same points.  The distances between Ir$^{4+}$
 and O$^{2-}$ order this way: $|A5|=|A6|=|B5^{\prime}|=|B6^{\prime}|
 >|A3|=|A4|=|B3^{\prime}|=|B4^{\prime}|>|A1|=|A2|=|B1^{\prime}|=|B2^{\prime}|$.
 The C$_2$ axis (thick dash line) orients along $\frac{1}{\sqrt{2}}
 (1,-1,0)$ at Ir$^{4+}$ A and $\frac{1}{\sqrt{2}}(0,1,1)$ at Ir$^{4+}$ B.
 Mapped to the ideal \hk\ lattice, A and B correspond to point $4$ and $8$ 
 in Fig.~\ref{fig:dm}, respectively.}
\label{fig:octahedron}
\end{figure}

Before moving to the microscopic theory of spin-orbit interactions, we
need to understand the electron energy levels of the Ir$^{4+}$
ions. With coordinates taken from Table. I in
Ref.~\onlinecite{takagi:prl}, two Ir$^{4+}$ and their surrounding
O$^{2-}$ are drawn in Fig.~\ref{fig:octahedron}. For A ion, the C$_2$
axis orients along $\frac{1}{\sqrt{2}}(1,-1,0)$. Under this symmetry
operation, $x \rightarrow -y$, $y \rightarrow -x$ and $z \rightarrow
-z$. Accordingly, we can group the $5d$ orbitals into even and odd
parity sectors, as shown in Table.~\ref{tab:state}.

\begin{figure}[hbt]
	\centering
   \includegraphics[width=3.0in]{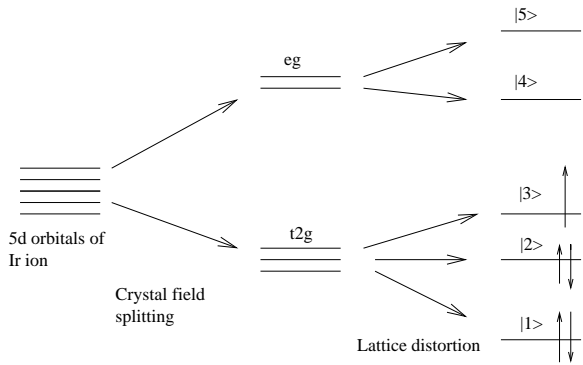}
 \caption{The splitting and electron occupation of $5d$ orbitals of Ir$^{4+}$ ions
  in the absence of spin-orbit interaction. The states are defined in 
  Table.~\ref{tab:state}.}
 \label{fig:sp}
\end{figure}

A large cubic crystal field splits the $e_g$ and $t_{2g}$ states.  The
surrounding O$^{2-}$ octahedron is slightly distorted to further split
all the three $t_{2g}$ states.  Ultimately no degeneracy is protected
because the C$_2$ symmetry has only one dimensional irreducible
representations.  The energetic ordering of orbitals shown in
Fig.~\ref{fig:sp} was determined by looking at Coulomb interaction 
from surrounding O$^{2-}$ and ignoring the spin-orbit interaction.  

\begin{table}
\begin{tabular}{|l|l|l|l|} 
\hline 
state               & $5$d orbitals at A          & $5$d orbitals at B             & parity 
\\ \hline
$\mid 1\rangle$     & $xy$                        & $yz$                           & even
\\ \hline
$\mid 2\rangle$     & $\frac{1}{\sqrt{2}}(xz-yz)$ & $\frac{1}{\sqrt{2}}(yx+zx)$    & odd
\\ \hline
$\mid 3\rangle$     & $\frac{1}{\sqrt{2}}(xz+yz)$ & $\frac{1}{\sqrt{2}}(yx-zx)$    & even
\\ \hline
$\mid 4\rangle$     & $x^2-y^2$                   & $y^2-z^2$                      & odd
\\ \hline
$\mid 5\rangle$     & $3z^2-r^2$                  & $3x^2-r^2$                     & even
\\ \hline
\end{tabular}
\caption{The parity sectors of $5d$ electron orbitals by C$_2$ rotation}
\label{tab:state}
\end{table}

\subsection{Microscopic theory of exchange spin Hamiltonian}
\label{sec:micro}

Though symmetry determines the allowed non-zero components of the DM
interaction, it does not give any guidance as to their relative and
absolute magnitudes.\cite{moriya:pr,elhajal:prb,koshibae:prb} In this
part, we will derive the exchange spin Hamiltonian from a microscopic
point of view and obtain expressions from which crude estimates of the
magnitude of various terms can be obtained\cite{moriya:pr,elhajal:prb,
  koshibae:prb} We consider both the hopping between Ir and O orbitals,
and direct hopping between Ir orbitals. We also assume that the
e$_g$-t$_{2g}$ splitting is much greater than the splittings among the
three t$_{2g}$ states so that we can completely project out the two
e$_g$ states.  The model is then of five electrons on on the $t_{2g}$
orbitals of every Ir$^{4+}$. Following some notations in
Ref.~\onlinecite{koshibae:prb}, we can write the Hamiltonian of the Ir
and O sublattice as
\begin{equation}
{\mathcal H} = {\mathcal H}_0 + {\mathcal H}_t + {\mathcal H}_{LS}
\;, 
\label{eq:mH}
\end{equation}
where, 
\begin{eqnarray}
{\mathcal H}_0 
      &=&  \sum_{jm\sigma} \epsilon_m d^{\dagger}_{jm \sigma} d_{jm \sigma} 
           + \sum _{kn \sigma} \epsilon_{p_n} p^{\dagger}_{kn \sigma} p_{kn \sigma} 
      \nonumber \\
      &+&  \frac{U_d}{2} \sum_{jm m^{\prime} \sigma \sigma^{\prime}} d^{\dagger}_{jm \sigma}
           d^{\dagger}_{jm^{\prime}\sigma^{\prime}} d_{jm^{\prime}\sigma^{\prime}} 
           d_{jm \sigma}  
      \nonumber \\
      &+&  \frac{U_p}{2} \sum_{kn n^{\prime} \sigma \sigma^{\prime}} p^{\dagger}_{kn \sigma}  
           p^{\dagger}_{k n^{\prime} \sigma^{\prime}} p_{k n^{\prime} \sigma^{\prime}}
           p_{kn \sigma}  ,
       \\
{\mathcal H}_t 
      &=& \sum_{jm \sigma} \sum_{k(j)n} (t_{jm,kn} d^{\dagger}_{jm \sigma} 
          p_{kn \sigma} + H.c.) \nonumber
      \\
&& + \sum_{\langle jj'\rangle}\sum_{mm'} t^d_{jm,j'm'}
d_{jm\sigma}^\dagger d_{j'm'\sigma}^{\vphantom\dagger}, \\
{\mathcal H}_{LS} 
      &=&  \lambda \sum_j {\boldsymbol \ell}_j \cdot {\boldsymbol s}_j 
      \;.
\label{eq:mH0tls}
\end{eqnarray}
$k(j)$ denotes the O$^{2-}$ of the neighboring Ir$^{4+}$ site $j$,
$d^{\dagger}_{jm \sigma}$ is the creation operator of an electron with
spin $\sigma$ of the $m$th $5d$ orbital of $i$th Ir ion, $\epsilon_m$ is
the energy of this orbital. $m$ will take $1,2,3$. $p^{\dagger}_{kn
  \sigma}$ is the creation operator of an electron on the $2p_n$ orbital
with spin $\sigma$.  The energies are measured from the lowest energy
level of the Ir $5d$ orbitals, and $U_d$ and $U_p$ are the Coulomb
interaction constants between holes on the Ir$^{4+}$ site and O$^{2-}$
site, respectively. We assume that $U_d$ and $U_p$ are
orbital-independent and ingore other ``Kanamori
parameters'':\cite{kanamori:1963} the inter-orbital exchange coupling
and the pair-hopping amplitude, which should be small compared with
Coulomb interaction. We also ignore the Coulomb interaction between two
eletrons on different intermediate O$^{2-}$ ions. Here $t_{jm,kn}$
denotes the transfer of an electron between the $m$th orbital of
Ir$^{4+}$ ion $j$ and one of the $2p_n$ orbitals of the neighboring
O$^{2-}$ ions $k$. Similarly, $t^d_{jm,j'm'}$ is the matrix element for
electron transfer between $m$ and $m'$ orbitals on two nearest-neighbor
(in the \hk\ sense) Ir atoms.  ${\boldsymbol \ell}_j$ and ${\bf s}_j$
denote the orbital and spin angular momenta at the $j$th Ir$^{4+}$ ion,
respectively, and $\lambda$ is the spin-orbit coupling constant of the
Ir$^{4+}$ ion.

In order to understand the electron occupation on each site,
we collect the quadratic terms for each site in Eq.~\eqref{eq:mH0tls} and
write down the onsite Hamiltonian as
\begin{equation}
{\mathcal H}^{(i)} 
               = \sum_{mm^{\prime}\sigma\sigma^{\prime}}
                 d_{im\sigma}^{\dagger} {\mathcal M}^{(i)}_{m\sigma,m^{\prime}\sigma^{\prime}}
                 d_{im^{\prime}\sigma^{\prime}}
\label{eq:onsite}                 
\end{equation}
with
\begin{equation}
{\mathcal M}^{(i)}_{m\sigma,m^{\prime}\sigma^{\prime}} 
       = {\epsilon}_m \delta_{\sigma\sigma^{\prime}}\delta_{mm^{\prime}}
         + \lambda{\boldsymbol \ell}_{imm^{\prime}} \cdot  
           {\boldsymbol{s}}_{\sigma \sigma^{\prime}}
         \; ,
\label{eq:matrixelement}
\end{equation}                     
where $\boldsymbol{\sigma}$ is the Pauli matrix and ${\bf L}_{imm^{\prime}}$ is 
the matrix element of ${\bf L}_i$
between the $m$th and $m^{\prime}$th orbital of the $i$th Ir$^{4+}$
ion.  It is useful to note\cite{goodenough:pr} that the vector of three-dimensional matrix  orbital
angular momentum operators projected into the $t_{2g}$ manifold is
actually proportional to the vector of orbital angular momentum
operators for the 3 ordinary ($p_x,p_y,p_z$) states, but with a
proportionality constant of $-1$!  That is, suppressing the $m,m'$
indices,
\begin{equation}
  \label{eq:21}
  {\boldsymbol \ell}_i = - {\boldsymbol{\sf L}}_i,
\end{equation}
where ${\boldsymbol{\sf L}}$ is a canonical angular momentum operator
with ${\sf L}^2 = \ell(\ell+1) = 2$.  This effectively makes the
spin-orbit coupling term directly analogous to the familiar one from an
isolated atom with spherical symmetry in a $p$ shell, but with the sign
of the spin-orbit coupling reversed.

\subsection{Strong and weak spin orbit limits}
\label{sec:strong-weak-spin}

Obviously the nature of the ``spin'' itself (i.e. the Kramer's doublet
ground state of the single whole in this multiplet) is crucially
dependent upon the strength of the spin-orbit interaction, relative to
the non-cubic splittings $\epsilon_3-\epsilon_2,
\epsilon_3-\epsilon_1$.  This determines the nature of the {\sl
  wavefunctions} of the Kramer's pair, for instance the degree to which
the ``spin''  carries true electron spin angular moment or instead
orbital angular momentum.  This is more fundamental than
the exchange interaction, so we consider it first.  

\subsubsection{Strong spin orbit}
\label{sec:strong-spin-orbit}

In the strong spin-orbit limit, we can to a first approximation ignore
the non-cubic splittings, and we have simply 
\begin{equation}
  \label{eq:22}
  {\mathcal M}^{(i)} = \lambda {\boldsymbol\ell}_i\cdot {\bf s}_i = -
  \lambda {\boldsymbol{\sf L}}_i \cdot {\bf s}_i.
\end{equation}
This is of course diagonalized by constructing eigenstates of the
``total angular momentum'' 
\begin{equation}
  \label{eq:23}
  {\bf J}_i = {\boldsymbol {\sf L}}_i + {\bf s}_i.
\end{equation}
Because of the minus sign in Eq.~(\ref{eq:22}), the highest energy
doublet is simply the $j=1/2$ Kramer's pair.  This describes the
wavefunction of the half-filled orbital.  It is natural to define the
effective spin operator in this case as
\begin{equation}
  \label{eq:24}
  {\bf S}_i = {\bf J}_i.
\end{equation}
 Clearly it is a strong mix of orbital and
spin components.  According to the Wigner-Eckart  theorem, the matrix
elements of ${\bf s}_i$, ${\boldsymbol{\sf L}}_i$ and ${\bf J}_i$ are
all proportional.  This enables one, with a little Clebsch-Gordan
algebra, to arrive at an expression for the magnetic moment operator (in
the $j=1/2$ manifold)
\begin{equation}
  \label{eq:25}
  {\bf M}_i = -\mu_B ({\boldsymbol\ell}_i+2{\bf s}_i) = +2 \mu_B {\bf S}_i,
\end{equation}
where $\mu_B$ is the Bohr magneton.  Interestingly, this is the same
magnitude but opposite sign as for a free electron!  It will of course
suffer corrections perturbative in $(\epsilon_i-\epsilon_3)/\lambda$, as
one moves away from the strong spin orbit limit.

\subsubsection{Weak spin orbit}
\label{sec:weak-spin-orbit}

Now consider the weak spin orbit limit.  In this case, for $\lambda=0$,
the half-filled doublet is simply the $m=3$ orbital, with two possible
``true'' spin orientations.  Thus we approximately have 
\begin{equation}
  \label{eq:26}
  {\bf S}_i \approx {\bf s}_i + \mathcal{O}(\lambda/(\epsilon_i-\epsilon_j)).
\end{equation}
Now there is essentially no orbital angular momentum component to the
spin (${\boldsymbol\ell}_i \approx 0$), and one obtains 
\begin{equation}
  \label{eq:27}
 {\bf M}_i= 
   - 2 \mu_B {\bf S}_i (1+{\mathcal
     O}(\frac{\lambda}{|\epsilon_{1,2}-\epsilon_3|}))          \;.
\end{equation}
Note the important sign difference from Eq.~(\ref{eq:25}).  This is the
most fundamental physical distinction between the weak and strong
spin-orbit limits.  However, the magnitude of the proportionality
between the magnetization and spin -- the $g$-factor -- is the same in
both cases.  This means that the simplest experimental measure, the
Curie susceptibility, cannot distinguish the two possibilities.  We will
consider both cases below.  

\subsection{General exchange formulation}
\label{sec:gener-exch-form}

We now turn to the exchange calculations.  Let us consider the general
case first.  We must deal with ${\mathcal M}^{(i)}$, which is a
$6\times6$ matrix. Diagonalize ${\mathcal M}^{(i)}$ so that ${\mathcal
  M}^{(i)} = {T^{(i)}}^{\dagger} {\mathcal E} T^{(i)}$. Here, ${\mathcal
  E}$ is a site-independent eigenvalue matrix, and $T^{(i)}$ is a
unitary eigenvector matrix.  ${\mathcal M}^{(i)}$ has three different
eigenvalues ${\mathcal E}_1$, ${\mathcal E}_2$ and ${\mathcal E}_3$,
each has a two-fold degeneracy due to Kramers' degeneracy theorem.  The
effective spin operator ${\bf S}_i$ will be defined to act in this
doublet.  In the strong and weak spin-orbit limits, we have explicitly
Eq.~(\ref{eq:24}) and Eq.~(\ref{eq:26}), respectively.  Furthermore, we
define a new set of electron creation and annihilation operators
\begin{equation}
a_{im\sigma} = T^{(i)}_{m\sigma,m^{\prime}\sigma^{\prime}} d_{im^{\prime}\sigma^{\prime}}
\label{eq:xform}
\end{equation}
with $a_{im\sigma}$ annihilates an electron on the ${\mathcal E}_m$ state 
with spin $\sigma$ at site $i$. 

Without losing any generality, we assume that ${\mathcal E}_3 >
{\mathcal E}_{1,2}$, then ${\mathcal E}_{1,2}$ states are fully occupied
and ${\mathcal E}_3$ state is half-occupied, leading to a total
spin-$\frac{1}{2}$ at every site.  Accordingly, the magnetic momentum
operator (${\bf M}_i = - \mu_B ({\boldsymbol\ell}_i + 2 {\bf s}_i) $ at
each site should be projected onto the Kramers' doublet ground states:

\begin{eqnarray}
      \frac{{\bf M}_i}{ \mu_B}
      & = & -  P_i \sum_{mn \alpha \beta} 
                        d^{\dagger}_{im\alpha}
            ({\boldsymbol\ell}_{imn}\delta_{\alpha \beta}+\delta_{mn}\boldsymbol{\sigma}_{\alpha \beta})
                       d_{in\beta} P_i
        \nonumber  \\
      & = &       -{\bf G}^{(i)}_{3\alpha,3\beta}
      {\boldsymbol\sigma}_{\beta\alpha}\cdot {\bf S}_i
\label{eq:moment}
\end{eqnarray}
with $\boldsymbol\sigma$ the vector of Pauli matrices. Also, ${\bf G}^{(i)}_{l\sigma,j\delta}$ and 
the effective spin operator ${\bf S}_i$ are defined as 
\begin{eqnarray}
    {\bf G}^{(i)}_{l\sigma,j\delta}  
      & = &  \sum_{mn \alpha \beta} T^{(i)}_{l\sigma,m\alpha}
           ({\boldsymbol\ell}_{imn}\delta_{\alpha \beta}+\delta_{mn}\boldsymbol{\sigma}_{\alpha \beta})
             T^{(i)\ast}_{j\delta,n\beta} , \nonumber
                  \\
    {\bf S}_i
      & = &  \sum_{\alpha, \beta} \frac{1}{2} 
             a_{i3\alpha}^{\dagger} \boldsymbol{\sigma}_{\alpha \beta} a_{i3\beta} 
       \;,
\end{eqnarray}
and $P_i$ is the ground state projection operator:
\begin{equation}
   P_i =   a_{i3\uparrow}  \mid \phi \rangle \langle \phi \mid a_{i3\uparrow}^{\dagger}
         + a_{i3\downarrow}\mid \phi \rangle \langle \phi \mid a_{i3\downarrow}^{\dagger}
\;.
\end{equation}
Here $\mid \phi \rangle$ is the ${\mathcal E}_{1,2,3}$ fully-occupied state.
In the last step Eq.~\eqref{eq:moment}, $\sum_{\sigma} a^{\dagger}_{i3\sigma}a_{i3\sigma} = 1$
has been used.

Let's go back to Eq.~\eqref{eq:mH0tls}, and express the microscopic
Hamiltonian in terms of $a_{jm\sigma}$ and $a^{\dagger}_{jm\sigma}$.
Given the Hamiltonian in Eq.~(\ref{eq:mH}), which includes the largest
Coulomb energy $U$ but neglects the smaller Hunds-rule exchange coupling
{\sl between} electrons in different orbitals on the same atom (and
other similar interactions), only hopping through the half-filled
orbital contributes to the super-exchange interaction.  This is in
accord with the ``Goodenough-Kanamori'' rules, which state that the
exchange coupling contributed from a half-occupied orbital and a
fully-occupied orbitals is much weaker than the one from two
half-occupied orbitals.  Thus, we only need to focus on the hopping
between the ${\mathcal E}_3$ orbitals, as half-occupied orbital.  
The microscopic Hamiltonian is written as
\begin{widetext}
\begin{eqnarray}
  {\mathcal H} 
    &=&    \sum_{kn \sigma} \epsilon_{p_n} p^{\dagger}_{kn \sigma} 
           p_{kn \sigma} + \frac{U_p}{2} \sum_{kn n^{\prime} \sigma \sigma^{\prime}} p^{\dagger}_{kn \sigma}  
           p^{\dagger}_{k n^{\prime} \sigma^{\prime}} p_{k n^{\prime} \sigma^{\prime}}
           p_{kn \sigma}  
          +\sum_{jm\sigma}{\mathcal E}_m a^{\dagger}_{jm\sigma} a_{jm\sigma}
           +  \frac{U_d}{2} \sum_{jm m^{\prime} \sigma \sigma^{\prime}} a^{\dagger}_{jm \sigma}
           a^{\dagger}_{jm^{\prime}\sigma^{\prime}} a_{jm^{\prime}\sigma^{\prime}} 
           a_{jm \sigma}  
     \nonumber \\
    &+&    \sum_{jk(j)n}\sum_{\alpha \beta}[(\tilde{t}_{j3,kn}\delta_{\alpha \beta} 
          + {\bf C}_{j,kn} \cdot {\boldsymbol{\sigma}}_{\alpha \beta})
             a^{\dagger}_{j3\alpha} p_{kn\beta} + H.c.]+ \sum_{\langle
               jj'\rangle} \sum_{\alpha\beta} [ \left(
               \tilde{t}^d_{j3,j'3} \delta_{\alpha\beta} + {\bf
                 C}^d_{jj'}\cdot{\boldsymbol\sigma}_{\alpha\beta}\right)
           a_{j3\alpha}^\dagger a_{j'3\beta}^{\vphantom\dagger} + H.c. ]
     \;,   
\label{eq:effh}
\end{eqnarray}
with 
\begin{eqnarray}
    \tilde{t}_{j3,kn} &=& \sum_{m\sigma} \frac{1}{2}t_{jm,kn}T^{(j)}_{3\sigma,m\sigma}
    \nonumber \\
    {\bf C}_{j,kn} &=& \sum_{m,\alpha \beta} \frac{1}{2} t_{jm,kn}
    T^{(j)}_{3\alpha,m\beta} {\boldsymbol \sigma}_{\beta\alpha}
                \;,
\label{eq:para}
\end{eqnarray}
and
\begin{eqnarray}
\tilde{t}^d_{j3,j'3} &=& \sum_{mm^{\prime},\alpha \sigma} \frac{1}{2} t^d_{jm,j'm'} 
                  T^{(j)}_{3\alpha,m\sigma} {T^{(j')}}^{\dagger}_{m'\sigma,3\alpha}
                                   \nonumber \\
{\bf C}^d_{jj'} & = & \sum_{mm^{\prime},\sigma,\alpha \beta} \frac{1}{2} t^d_{jm,j'm'}
                T^{(j)}_{3\alpha,m\sigma} {T^{(j')}}^{\dagger}_{m'\sigma,3\beta}
                {\boldsymbol \sigma}_{\beta\alpha}
                \;,
\end{eqnarray}
where $\boldsymbol{\sigma}$ is vector of the three Pauli matrices.
Now we may follow the standard perturbative treatment of superexchange.
We consider separately the superexchange through the intermediate
O$^{2-}$ ions, and the direct exchange contributions.

\subsubsection{Superexchange through oxygen ions}
\label{sec:super-thro-o2}

In this case the leading contribution is fourth order in hopping, i.e. a
result of fourth order degenerate perturbation  theory.  We must include four
``hops'' between Ir$^{4+}$ and O$^{2-}$ ions, which consist of ``hops''
described by spin-isotropic $\tilde{t}$ matrix elements, and ``hops''
given by anisotropic ${\bf C}$ matrix elements.
One thereby obtains the exchange Hamiltonian as 
\begin{equation}
 {\mathcal H}_{ex} 
       = \sum_{\langle i j \rangle} [ J {\bf S}_i \cdot {\bf S}_j +
         {\bf D}_{ij} \cdot ({\bf S}_i \times {\bf S}_j) + {\bf S}_i \cdot
         \overleftrightarrow{\Gamma}_{ij} \cdot {\bf S}_j ]
\label{eq:ex}
\end{equation}
with the first two terms the Heisenberg and DM
interactions precisely as in Eq.~(\ref{eq:hamiltonian}), and the third 
term the anisotropic exchange. The explicit formulae for the coupling 
constants are:
\begin{eqnarray}
         J &=& 4 \sum_{kn,k^{\prime}n^{\prime}} s_{ij,kn}g_{kn,k^{\prime}n^{\prime}}s_{ji,k^{\prime}n^{\prime}}
       \label{eq:18}\\
         {\bf D}_{ij} &=&
             - 4i \sum_{kn,k^{\prime}n^{\prime}} ({\bf v}_{ij,kn}g_{kn,k^{\prime}n^{\prime}}s_{ji,k^{\prime}n^{\prime}}
             - s_{ij,kn} g_{kn,k^{\prime}n^{\prime}} {\bf v}_{ji,k^{\prime}n^{\prime}})
         \\
         \overleftrightarrow{\Gamma}_{ij} &=& 
          4 \sum_{kn,k^{\prime}n^{\prime}} [(\overleftarrow{\bf v}_{ij,kn}g_{kn,k^{\prime}n^{\prime}}
          \overrightarrow{\bf v}_{ji,k^{\prime}n^{\prime}} +\overleftarrow{\bf v}_{ji,kn} g_{kn,k^{\prime}n^{\prime}}
          \overrightarrow{\bf v}_{ij,k^{\prime}n^{\prime}})- \overleftrightarrow{1} 
          ({\bf v}_{ij,kn} \cdot g_{kn,k^{\prime}n^{\prime}} {\bf v}_{ji,k^{\prime}n^{\prime}})]
     \;.
\label{eq:jdm}
\end{eqnarray}
The vector with arrow $\leftarrow$ or $\rightarrow$ indicates that inner product
is taken with the spin operator put in the direction of the arrow. 
$\overleftrightarrow{1}$ is a $3 \times 3$ unit matrix. $s_{ij,kn}$, ${\bf v}_{ij,kn}$ and
$g_{kn,k^{\prime}n^{\prime}}$ are given by
\begin{eqnarray}
  s_{ij,kn} &=&  \tilde{t}_{i3,kn} \tilde{t}_{kn,j3} + 
                 {\bf C}_{i,kn} \cdot {\bf C}_{kn,j}
           \\
  {\bf v}_{ij,kn} &=&   {\bf C}_{i,kn}\tilde{t}_{kn,j3}
      + \tilde{t}_{i3,kn}{\bf C}_{kn,j}+i ({\bf C}_{i,kn} \times {\bf C}_{kn,j})
          \\            
  g_{kn,k^{\prime}n^{\prime}} 
      &=&   \frac{(1-\frac{1}{2}\delta_{kk^{\prime}}\delta_{nn^{\prime}})
            (\tilde{\epsilon}_{p_{kn}}^{-1}+\tilde{\epsilon}_{p_{k^{\prime}n^{\prime}}}^{-1})^2}
            {\tilde{\epsilon}_{p_{kn}} + \tilde{\epsilon}_{p_{k^{\prime}n^{\prime}}} + U_p   
            \delta_{kk^{\prime}}}
             + (\tilde{\epsilon}_{p_{kn}} \tilde{\epsilon}_{p_{k^{\prime}n^{\prime}}} U_d)^{-1}
\label{eq:g}
\end{eqnarray}
with $\tilde{\epsilon}_{p_{kn}}= {\mathcal E}_3 - \epsilon_{p_{kn}}+5(U_d-U_p)$.
In the following subsections, we will try to estimate these exchange couplings
in both the strong and weak spin-orbit interaction cases.

\subsubsection{Direct exchange}
\label{sec:direct-exchange}

Here we require only second order perturbation theory in the direct
matrix elements.  One obtains the results\cite{moriya:pr}:
\begin{eqnarray}
   J &=& \frac{2 {\left| \tilde{t}^d_{ij} \right| }^2}{U_d},
           \\
{\bf D}_{ij} &=& -\frac{4i}{U_d}( {\bf C}^d_{ij} \tilde{t}^d_{ji} - \tilde{t}^d_{ij}{\bf C}^d_{ji} ),
           \\
\overleftrightarrow{\Gamma}_{ij} &=& \frac{4}{U_d}(\overleftarrow{\bf C}^d_{ij} \overrightarrow{\bf C}^d_{ji} 
                            + \overleftarrow{\bf C}^d_{ji} \overrightarrow{\bf C}^d_{ij} - 
                            \!\overleftrightarrow{1}\! ({\bf C}^d_{ij} \cdot {\bf C}^d_{ji}))
                          \;.
\end{eqnarray}
\end{widetext}

\subsection{Strong spin-orbit interaction}
\label{sec:strong}

As discussed above in Sec.~\ref{sec:strong-spin-orbit}, in the strong
spin-orbit limit, $\lambda \gg |\epsilon_{1,2} - \epsilon_3|$, one can
obtain effective total angular momentum eigenstates with $j=1/2$.
Choosing Eq.~(\ref{eq:24}), and rewriting the corresponding eigenstates
in the canonical $t_{2g}$ basis, Eq.~\eqref{eq:xform} becomes
\begin{eqnarray}
a_{i3\uparrow} &=&   \frac{1}{\sqrt{3}} 
                     ((-i) d_{i,xz \downarrow} + d_{i,yz\downarrow}
                      +  d_{i,xy\uparrow})
           \\
a_{i3\downarrow} &=&  \frac{1}{\sqrt{3}} 
                      ((i) d_{i,xz \uparrow} + d_{i,yz\uparrow}
                      - d_{i,xy\downarrow})
                      \;,
\label{eq:xformstrong}
\end{eqnarray} 
in which, we have expressed $a_{i3\uparrow}$/$a_{i3\downarrow}$ in terms
of the $t_{2g}$ annihilation operator to avoid the position dependence
of the coefficients. 

\subsubsection{Superexchange through oxygen ions}
\label{sec:super-thro-o2-1}

The complicated expression of Eq.~\eqref{eq:jdm} requires simplification
if we want to have a quantative understanding of the exchange coupling.
However, some information can be immediately obtained from
Eq.~\eqref{eq:xformstrong}, in particular that all
$\tilde{t}_{i3,kn}=0$, which makes $J$, ${\bf D}_{ij}$ and
$\overleftrightarrow{\Gamma}_{ij}$ only remain terms with ${\bf
  C}_{i,kn}$. To simplify further, we need some explicit form for the
transfer integrals $t_{jm,kn}$.  Hence, we will make further
approximation that the surrounding octahedra of Ir$^{4+}$ are perfect so
that we can apply the cubic symmetry to find out the nonvanishing
transfer integrals and also the relation between them, which is listed
in Table.~\ref{tab:integral} for Ir$^{4+}$ A and B in
Fig.~\ref{fig:octahedron}.  Deviations from these forms should
presumably be small, since the non-cubic distortion is.
\begin{table}
\begin{tabular}{|l|l|l|l|l|l|l|} 
\hline 
            & $2p_x$  & $2p_y$  & $2p_z$ & $5p_x$ & $5p_y$ & $5p_z$          
\\ \hline
A, $xz$     & $t$     & $0$     & $0$    & $0$    & $0$    & $0$ 
\\ \hline
A, $yz$     & $0$     & $t$     & $0$    & $0$    & $0$    & $-t$ 
\\ \hline
A, $xy$     & $0$     & $0$     & $0$    & $-t$   & $0$    & $0$ 
\\ \hline
B, $xz$     & $0$     & $0$     & $0$    & $-t$   & $0$    & $0$ 
\\ \hline
B, $yz$     & $0$     & $0$     & $t$    & $0$    & $-t$   & $0$     
\\ \hline
B, $xy$     & $t$     & $0$     & $0$    & $0$    & $0$    & $0$ 
\\ \hline
\end{tabular}
\caption{The transfer integrals between the $t_2g$ orbitals on A and B
Ir$^{4+}$ and the $p_{x,y,z}$ orbitals on the intermediate O$^{2-}$ ions.
``$2p_x$'' represents the $p_x$ orbital on the $2$nd O$^{2-}$ ion in Fig.~\ref{fig:octahedron},
``A, $xz$'' represents the $xz$ orbital on the A ion, And the entry $t$ on the row of 
``A, $xz$'' and the column of ``$2p_x$'' denotes the hopping amplitude (transfer integral) 
from $xz$ orbital at A ion to $p_x$ orbital on $2$nd O$^{2-}$ ion. Other notation can be 
understood likewise.}
\label{tab:integral}
\end{table}

Based on the transfer integrals listed in Table.~\ref{tab:integral}, we
evaluate the exchange coupling constant $J$ and
$\overleftrightarrow{\Gamma}_{AB}$.  For bond AB, collecting non-zero
coupling constants (actually $J=0$, ${\bf D}_{AB} =0$), we obtain
\begin{equation}
{\mathcal H}_{AB} = - J S_A^x S_B^x + J S_A^y S_B^y + J S_A^z S_B^z
\label{eq:ab}
\end{equation}
with 
\begin{equation}
J = \frac{4}{9} |t|^4 (2g_{2p_x,5p_x} - g_{2p_x,2p_x} - g_{5p_x,5p_x})
\;.
\end{equation}
Since from Eq.~\eqref{eq:g} $g_{2p_x,5p_x}>g_{2p_x,2p_x},g_{5p_x,5p_x}$,
then $J>0$.  Thus we find ferromagnetic interaction between the x
components and antiferromagnetic interactions between th y and z
components along this link.  This corresponds to the form in
Eq.~(\ref{eq:1}) of the Introduction, with
$\epsilon_{ij}^y=\epsilon_{ij}^z=-\epsilon_{ij}^x=1$ for this link.  

Because all links are equivalent by point group operations, we can
deduce the exchange interactions of all other bonds by symmetry.  The
sites A and B correspond to point $4$ and $8$ in our notation in
Fig.~\ref{fig:dm}.  The result is that the exchange interactions on each
bond are ferromagnetic between one component, and antiferromagnetic
between the other two.  These principle components are always along $x$,
$y$, or $z$.  We will call a bond in which the $x$ component is
ferromagnetic a ``type $x$ bond'', and similarly for $y$, $z$.  The type
of each bond is listed in Table.~\ref{tab:hamtab}. This Hamiltonian
breaks spin-rotational symmetry strongly.  A simple rule can be used
to characterize the Hamiltonian of a given bond: if bond $(ij)$ is
located in y-z plane, then the bond is type $x$ bond and has type $x$
exchange Hamiltonian; if it is located in x-z plane, then the bond is
type $y$ bond and has type $y$ exchange Hamiltonian; if it is located in
x-y plane, the bond is type $z$ bond and has type $z$ exchange
Hamiltonian. As a result, the three bonds in every triangle (See
Fig.~\ref{fig:dm}) have different exchange Hamiltonian. The ground
states of this Hamiltonian will be studied in Sec.~\ref{sec:strongspin}.

\begin{table}
\begin{tabular}{|l|l|l|} 
\hline 
             type $x$    & type $y$    & type $z$        
\\ \hline
             $(1,2)$     & $(1,3)$     & $(2,3)$    
\\ \hline
             $(3,5)$     & $(3,4)$     & $(4,5)$    
\\ \hline
             $(\overline{5},7)$     & $(\overline{5},6)$     & $(6,7)$    
\\ \hline
             $(4,8)$     & $(8,9)$     & $(4,9)$    
\\ \hline
             $(8,\overline{11})$    & $(7,\overline{11})$    & $(7,8)$        
\\ \hline
             $(\overline{1},\overline{6})$     & $(\overline{6},12)$    & $(\overline{1},12)$   
\\ \hline
             $(9,10)$    & $(\overline{2},9)$     & $(\overline{2},10)$  
\\ \hline
             $(10,12)$   & $(10,11)$   & $(11,12)$   
\\ \hline
\end{tabular}
\caption{The bond types of $24$ bonds in one unit cell. Points and bonds are based on
the notation in Fig.~\ref{fig:dm}. ``$\overline{i}$'' is used for the points which are 
simply a translation by a basis vector from point ``$i$''.}
\label{tab:hamtab}
\end{table}

\subsubsection{Direct exchange}
\label{sec:direct-exchange-1}

We consider two Ir atoms A and B, connected by a line
along the $(0,1,-1)$ direction.  There are two principle overlaps.  The
largest, whose magnitude we denote $t_1^d$, is between the $yz$ orbitals
at each atom -- this is a $\sigma$-bond.  A secondary overlap, of
magnitude $t_2^d$,  occurs
between orbitals of the form $xy-xz$ at each site, which corresponds to
$\pi$ bonding.  All other overlaps are expected to be negligible or
zero.  This leads remarkably to
\begin{eqnarray}
  \label{eq:28}
  \tilde{t}^d_{j3,j'3} & = & (t_1^d+t_2^d)/3, \qquad {\bf C}_{jj'}=0.
\end{eqnarray}
The result appears isotropic, despite the strong spin-orbit
interactions!  As a consequence, one obtains only Heisenberg exchange,
and ${\bf D}_{ij} =  \overleftrightarrow{\Gamma} = 0$!  It is remarkable
that one finds apparent isotropy even though the spin itself contains a
substantial orbital component.  As seen from the superexchange
calculation above, this is by no means guaranteed.  

The first corrections to the
strong spin orbit limit are linear in the non-cubic splittings, and
produce corrections to the Heisenberg model.  This occurs by a
contribution to ${\bf C}_{jj'}$ of ${\mathcal
  O}(|\epsilon_3-\epsilon_{1,2}|/\lambda)$.  The leading spin-orbit
corrections to the exchange Hamiltonian are then of the DM form, and
constrained by symmetry according to considerations of
Sec.~\ref{sec:symmetry-allowed-dm}.

\subsection{Weak spin-orbit interaction}
\label{sec:estimates}

In this part, we are going to look at the weak spin-orbit interaction
limit, $\lambda \ll
|\epsilon_1-\epsilon_2|,|\epsilon_2-\epsilon_3|$. This is the regime
which was often studied in
literature.\cite{moriya:pr,elhajal:prb,koshibae:prb} Standard
perturbation treatment can be applied, which yields
\begin{equation}
a_{jm\sigma} = d_{jm\sigma} + \frac{\lambda}{2}\sum_{m^{\prime} \sigma^{\prime}} 
     \frac{{\boldsymbol\ell}_{imm^{\prime}}
     \cdot \boldsymbol{\sigma}_{\sigma \sigma^{\prime}}}{\epsilon_m-\epsilon_{m^{\prime}}}
     d_{j m^{\prime} \sigma^{\prime}}
\end{equation}
with ${\boldsymbol\ell}_{imm^{\prime}}$ introduced previously in 
Eq.~\eqref{eq:matrixelement}.  Using this in Eq.~\eqref{eq:moment}
reproduces Eq.~(\ref{eq:27}).

Keeping the exchange coupling constant to the linear order of
$\frac{\lambda}{|\epsilon_{1,2}-\epsilon_3|}$, we can ignore
$\overleftrightarrow{\Gamma}_{ij}$, as it is of ${\mathcal
  O}((\frac{\lambda}{|\epsilon_{1,2}-\epsilon_3|})^2)$ compared with
$J$, thus we only need to evaluate $J$ and ${\bf D}_{ij}$.

\subsubsection{Superexchange through oxygen ions}
\label{sec:super-thro-o2-2}

Since all the bonds and sites are equivalent, we can take
$ij$ as bond BA in Fig.~\ref{fig:octahedron}. Denote the unit
directional vectors for $D_1$, $D_2$ and $D_3$ as ${\bf e}_1$, ${\bf
  e}_2$ and ${\bf e}_3$. Ignoring the small effect of lattice distortion
on these vectors and taking the corresponding values for an ideal \hk\
lattice, we will get ${\bf e}_1=\frac{1}{\sqrt{2}}(0,-1,1)$, ${\bf
  e}_2=\frac{1}{\sqrt{3}}(-1,1,1)$ and ${\bf
  e}_3=-\frac{1}{\sqrt{6}}(2,1,1)$. 
Making the same approximation as in previous section, we can evaluate the
exchange coupling constants:
\begin{eqnarray}
    J   & = & |t|^4 g_{2p_x,2p_x}   \nonumber  \\
    D_1 & = & {\bf D}_{BA}\cdot {\bf e}_1 = \frac{\lambda}{\sqrt{2}} |t|^4 
              (\frac{g_{2p_x,2p_x}}{\epsilon_2-\epsilon_3} 
              -\frac{g_{5p_x,2p_x}}{\epsilon_1-\epsilon_3}) 
              \nonumber \\
    D_2 & = & {\bf D}_{BA}\cdot {\bf e}_2 = \frac{2\lambda}{\sqrt{3}}|t|^4
              \frac{g_{5p_x,2p_x}}{\epsilon_1-\epsilon_3}
              \nonumber \\
    D_3 & = & {\bf D}_{BA}\cdot {\bf e}_3 = -\frac{\lambda}{\sqrt{6}} |t|^4
              (\frac{3g_{2p_x,2p_x}}{\epsilon_2-\epsilon_3}
              -\frac{g_{5p_x,2p_x}}{\epsilon_1-\epsilon_3})
            \;.
  \label{eq:dmestimate}
\end{eqnarray}
The three DM components
we obtained in Eq.~\eqref{eq:dmestimate} are not independent from each other.
That's because we render some symmetry to the system by the transfer integrals.
Hence, we will still consider all three components to be independent.
As discussed in Sec.~\ref{sec:locenergy}, $\epsilon_3>\epsilon_2>\epsilon_1$, 
additionally, we have $g_{kn,k^{\prime}n^{\prime}} >0$, then we can confer
from Eq.~\eqref{eq:dmestimate} that $J>0$, $D_2<0$, $D_3$ is probably positive
due to a factor of $3$ in front of $g_{2p_x,2p_x}$ and the smaller denominator 
of the positive term than the negative term, and $|D_1|$ is probably small
compared to $|D_2|$ due 
to the cancellation of positive and negative terms.

Using Eqs.~\eqref{eq:dmestimate} and ignoring its specific expression, 
we may estimate the strength of DM interactions crudely. Then, we estimate crudely
\begin{equation}
  \label{eq:19}
  |D_i|/J \approx \lambda/|\epsilon_{1,2}-\epsilon_3|.  
\end{equation}
Since we assume $\lambda \ll |\epsilon_{1,2}-\epsilon_3|$; otherwise the
perturbative treatment doesn't holds.  We estimate the spin-orbit
coupling $\lambda \approx 0.4$eV, taken from
Refs.\onlinecite{vugman:rbf,murov:handbook} (although the reference is
not directly relevant to \nio, we can use their spin-orbit coupling as
an approximation).  The splitting of the $t_{2g}$ states due to the
non-cubic environment which determines $\epsilon_{1,2}-\epsilon_3$ is
difficult to estimate.  As mentioned in Refs.\onlinecite{murugavel:07},
the $e_g$-$t_{2g}$ splitting for [Ir(NH$_3$)$_6$]$^{3+}$ is about $5$eV.
However, if we seek a {\sl lower} bound on $|D_i|$ we can make due with
what is probably an over-estimate of this splitting.  Taking
$|\epsilon_{1,2}-\epsilon_3| \sim 5-10$eV is surely in that category,
and we therefore find $|D_i|/J \gtrsim 0.04-0.1$.

\subsubsection{Direct exchange}
\label{sec:direct-exchange-2}

One can similarly evaluate the induced DM terms at first order in the
spin-orbit coupling in the case of direct exchange.  One again obtains a
{\bf D}-vector consistent with the symmetry considerations in
Sec.~\ref{sec:symmetry-allowed-dm}.  

\section{Classical ground states of the strong exchange anisotropy
  Hamiltonian} 
\label{sec:strongspin}

In this Section, we will consider the ground states of the strongly
anisotropic Hamiltonian, Eq.~(\ref{eq:1}), obtained in the strong
spin-orbit limit from the Ir-O-Ir superexchange mechanism. 

Take the triangle $\Delta123$ in Fig.~\ref{fig:dm} for example. Bond
$(1,2)$ is of bond type $x$; bond $(1,3)$ is of bond type $y$; bond
$(2,3)$ is of bond type $z$.  Then for bond $(2,3)$, the Hamiltonian is
\begin{equation}
 {\mathcal H}_{(2,3)} = J(S_2^x S_3^x + S_2^y S_3^y - S_2^z S_3^z)
 \;.
\label{eq:eq49}
\end{equation}
Clearly ${\mathcal H}_{2,3}$ is minimized if
\begin{eqnarray} 
 {\bf S}_2^z &=& {\bf S}_3^z \\
 {\bf S}^{x,y}_2 &=& -{\bf S}^{x,y}_3 
 \;.
\label{eq:23z}
\end{eqnarray}
In general, for each bond, the energy is minimized if the
ferromagnetically interacting components of the two spins involved are
parallel, and the antiferromagnetically interacting components are
antiparallel.  We can search for {\sl unfrustrated} ground states by
demanding this on every bond.  Fixing one spin, its neighbors are
therefore determined, and from them further neighbors, etc.  It is
straightforward to verify that in this procedure {\sl no contradictions
  are encountered} despite the presence of loops on the lattice.  In
this way all classical ground states are determined from the choice of a
single initial spin.  Thus the Hamiltonian is unfrustrated, and we have
found its full set of classical ground states.  Mathematically, we can
write the full spin configuration as
\begin{equation}
{\bf S}_i =   s^x V_{1,i} + s^y V_{2,i}
            + s^z V_{3,i}
            \;,
\label{eq:eq123}
\end{equation}
where ${\bf s}=(s^x,s^y,s^z)$ is a unit vector, and $V_{a,i}$ is the
vector $V_a$ corresponding to the $i$th spin in
Table.~\ref{tab:strongbase}.  We see that the ground states are parametrized
by two continuous parameters -- the angles specifying the orientation of
the initial spin, or of ${\bf s}$.  This is actually an accidental degeneracy,
since the system has only discrete space-group symmetries, but it is
very small.  Still, it should be reduced to a discrete degeneracy by
perturbations such as quantum or thermal fluctuations, or additional
interactions, which will select a subset of these states.
\begin{table}

\begin{tabular}{|l|ccc|ccc|ccc|}\hline
           & \multicolumn{3}{|c|}{$V_1$} & \multicolumn{3}{|c|}{ $V_2$} & \multicolumn{3}{|c|}{$V_3$}
\\ \hline
Ir$^{4+}$  & $S_x$ & $S_y$ & $S_z$ & $S_x$ & $S_y$ & $S_z$ & $S_x$ & $S_y$ & $S_z$ 
\\ \hline
 $1$       & $-1$  & $0$   & $0$   & $0$   & $-1$   & $0$   & $0$   & $0$   & $1$
\\ \hline
 $2$       & $-1$  & $0$   & $0$   & $0$   & $1$   & $0$   & $0$   & $0$   & $-1$ 
\\ \hline
 $3$       & $1$   & $0$   & $0$   & $0$   & $-1$   & $0$   & $0$   & $0$   & $-1$
\\ \hline
 $4$       & $-1$  & $0$   & $0$   & $0$   & $-1$   & $0$   & $0$   & $0$   & $1$
\\ \hline
 $5$       & $1$   & $0$   & $0$   & $0$   & $1$   & $0$   & $0$   & $0$   & $1$
\\ \hline
 $6$       & $-1$  & $0$   & $0$   & $0$   & $1$   & $0$   & $0$   & $0$   & $-1$
\\ \hline
 $7$       & $1$   & $0$   & $0$   & $0$   & $-1$   & $0$   & $0$   & $0$   & $-1$
\\ \hline
 $8$       & $-1$  & $0$   & $0$   & $0$   & $1$   & $0$   & $0$   & $0$   & $-1$
\\ \hline
 $9$       & $1$   & $0$   & $0$   & $0$   & $1$   & $0$   & $0$   & $0$   & $1$
\\ \hline
 $10$      & $1$   & $0$   & $0$   & $0$   & $-1$   & $0$   & $0$   & $0$   & $-1$
\\ \hline
 $11$      & $-1$  & $0$   & $0$   & $0$   & $-1$   & $0$   & $0$   & $0$   & $1$
\\ \hline
 $12$      & $1$   & $0$   & $0$   & $0$   & $1$   & $0$   & $0$   & $0$   & $1$   
\\ \hline
\end{tabular}
\caption{The basis vectors for the ground state spin configurations of the strong
spin-orbit Hamiltonian.}
\label{tab:strongbase}
\end{table}

\section{Classical ground states induced by Dzyaloshinskii-Moriya interactions}
\label{sec:order}

\subsection{Order due to second component only}
  
In Sec.~\ref{sec:symm-micro}, we found that the direction of the DM vector for a
single bond is arbitrary, i.e. not determined from symmetry
considerations, and not calculable from microscopic theory without a
more detailed understanding of matrix elements than we have at present.
A general solution for the ground state with such an arbitrary DM vector
is quite difficult, because different triangles in the \hk\ lattice are
located in different planes.  In this subsection, we will consider the
special case in which the DM vector is normal to the triangular
plane, i.e. $D_1=D_3=0$ (see Fig.~\ref{fig:dm}).  This is a helpful
starting point for the more general case which we will address thereafter.

As in the case of the nearest-neighbor kagom\'{e} antiferromagnet, a
nonvanishing $D_2$ (here, by $D_2$ we mean the component of DM vector
which is normal to the kagom\'{e} plane) selects coplanar ground states
with $120^\circ$ spin orientations on each
triangle.\cite{elhajal:prb,ballou:pss,elhajal:pb} These are the only
configurations in which the Heisenberg interactions on a triangle are
minimized (i.e. the total sum of spins on a triangle is zero) {\sl and}
the DM interaction is minimized at the same time.  In the kagome
lattice, however, the coplanar ground state manifold is highly
degenerate, since rotating the spins in a single hexagon about the
normal axis of the kagom\'{e} plane by arbitrary angle generates a new
ground state from any other one. In contrast, for the \hk\ lattice, the
non-coplanar nature of different triangles reduces the degeneracy to
just a pair of Kramer's degenerate (reversed) states.  One of them is
drawn in Fig.~\ref{fig:hkorder1}; the other one is generated by
reversing all the spin directions.  The chirality of the \hk\ lattice
makes this state a ground state {\sl only} for $D_2<0$.  With the other
sign of $D_2$, the DM and Heisenberg interactions cannot be
simultaneously satisfied.  We will call these states uncanted ``windmill''
states -- see Fig.~\ref{fig:hkorder1}.

\begin{figure}
 \centering
 	\includegraphics[width=2.0in]{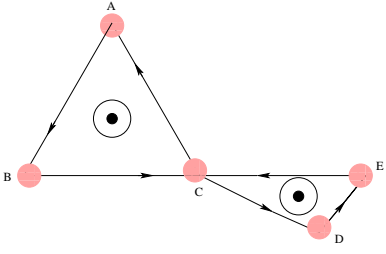}
	\caption{(Color online) Illustration of spin direction at shared corner 
	of two neighboring triangles. Black arrows indicate the DM interaction
	 path. B, C and E are in the same line.}
	\label{fig:orderexp}
\end{figure}

To see that the uncanted windmill states are the only classical ground
states, see Fig.~\ref{fig:orderexp}.  Starting from triangle ABS --
denoted $\Delta$ABC -- a nonvanishing $D_2$ component prefers a coplanar
spin configuration, which requires that spin A, B, C should lie in the
$\Delta$ABC plane and at 120 angles as dictated by the Heisenberg
interaction. The same applies to $\Delta$CDE. However, $\Delta$ABC and
$\Delta$CDE are not in the same plane, which confines the spin
orientation of site C to be aligned with the intersection line of
$\Delta$ABC and $\Delta$CDE. We apply this result to all spins, and the
Heisenberg interaction will select two states, which simultaneously
minimize the DM interaction with $D_2<0$. The magnetic unit cell of the
windmill state is the same as the chemical cell.

The result that $D_2<0$ completely removes the massive but accidental
ground state degeneracy of the \hk\ is quite dramatic.  A classical
antiferromagnet with this interaction will clearly order at low
temperature, and the drastic reduction in degeneracy suggests that even
for a quantum system, the suppression of quantum fluctuations by $D_2$
may be large.  Before turning to this, we continue with the analysis of
classical ordering in the remainder of this section.

\subsection{Magnetic representational analysis of space group}
\label{sec:representation}

Representational analysis of the magnetic space group has proven to be a
useful tool to extract information about low temperature ordered phases
using lattice
symmetry.\cite{wills:prb,wills:pb,dimmock:pr,bertaut:ac}. The idea is to
consider those types of magnetic order which can be reached by a
continuous transition from the paramagnetic state, which has the full
space group symmetry.  Though there is no a priori reason why the {\sl
ground state} configuration need be of this type, this is a convenient
way to generate candidate magnetically ordered states.  In principle,
one may iterate this procedure to generate lower temperature ordered
states, generating all possible ordered phases.

The operators of the space group act on both the position of the
magnetic ion and on the components of the spin vectors.  The combination
of these two results are described by the magnetic representation
$\Gamma$. The magnetic representation for a particular site can be
decomposed into contributions from the irreducible representations of
the little group
\begin{equation}
 \Gamma=\sum_{\mu}{n_{\mu}\Gamma_{\mu}}
 \;.
\end{equation}

For \nio, the space group is $P4_132$ (although it can also be $P4_332$,
the results should be equivalent),\cite{takagi:prl} and the Ir$^{4+}$
ions sit on the $12d$. Here, we only focus the simplest case when the
propogation vector $\vec{k}=(0,0,0)$. A program called
``SARAh''\cite{wills:pb} is used to do the decomposition of magnetic
representation
\begin{equation}
 \Gamma=1\Gamma_1^{(1)}+2\Gamma_2^{(1)}+3\Gamma_3^{(2)}+3\Gamma_4^{(3)}+3\Gamma_5^{(3)}
 \;,
\label{eq:dec}
\end{equation}
in which, the superindex represents the dimension of the irreducible
representations, and the subindex counts the irreducible representation.

Landau theory requires that only one representation can be involved in a
critical transition, and so with this constraint there are only five
possible magnetic structure for $\vec{k}=(0,0,0)$. Even within this
decomposition and Landau theory constraints, for certain representations
($\Gamma_3$,$\Gamma_4$,$\Gamma_5$), there still remain a lot of degrees
of freedom because of the multiple basis elements in these 3-dimensional
representations. For simplicity, we only discuss the one dimensional
representation $\Gamma_1$, $\Gamma_2$. The basis vectors for these two
representations calculated are given in Table.~\ref{tab:basis}.

\begin{table}

\begin{tabular}{|l|ccc|ccc|ccc|}\hline
& \multicolumn{3}{|c|}{$\Gamma_1^{(1)}$} & \multicolumn{6}{|c|}{ $\Gamma_2^{(1)}$ } 
\\ \hline
Basis vector & \multicolumn{3}{|c|}{$\psi_1$} & \multicolumn{3}{|c|}{ $\psi_2$} & \multicolumn{3}{|c|}{$\psi_3$}
\\ \hline
Ir$^{4+}$  & $S_x$ & $S_y$ & $S_z$ & $S_x$ & $S_y$ & $S_z$ & $S_x$ & $S_y$ & $S_z$ 
\\ \hline
 $1$ & $0$ & $\frac{1}{\sqrt{2}}$ & $\frac{1}{\sqrt{2}}$ & 0 & $\frac{1}{\sqrt{2}}$ & $-\frac{1}{\sqrt{2}}$ & $-1$ & $0$ & $0$
\\ \hline
 $2$ & $\frac{1}{\sqrt{2}}$ & $\frac{1}{\sqrt{2}}$ & $0$ & $\frac{1}{\sqrt{2}}$ & $-\frac{1}{\sqrt{2}}$ & $0$ & $0$ & $0$ & $-1$
\\ \hline
 $3$ & $\frac{1}{\sqrt{2}}$ & $0$ & $\frac{1}{\sqrt{2}}$ & $-\frac{1}{\sqrt{2}}$ & $0$ & $\frac{1}{\sqrt{2}}$ & $0$ & $-1$ & $0$
\\ \hline
 $4$ & $-\frac{1}{\sqrt{2}}$ & $\frac{1}{\sqrt{2}}$ & $0$ & $\frac{1}{\sqrt{2}}$ & $\frac{1}{\sqrt{2}}$ & $0$ & $0$ & $0$ &  $1$
\\ \hline
 $5$ & $0$ & $\frac{1}{\sqrt{2}}$ & $-\frac{1}{\sqrt{2}}$ & $0$ & $-\frac{1}{\sqrt{2}}$ & $-\frac{1}{\sqrt{2}}$ & $1$ & $0$ & $0$
\\ \hline
 $6$ & $-\frac{1}{\sqrt{2}}$ & $0$ & $\frac{1}{\sqrt{2}}$ & $\frac{1}{\sqrt{2}}$ & $0$ & $\frac{1}{\sqrt{2}}$ & $0$ & $1$ & $0$
\\ \hline
 $7$ & $\frac{1}{\sqrt{2}}$ & $\frac{1}{\sqrt{2}}$ & $0$ & $-\frac{1}{\sqrt{2}}$ & $\frac{1}{\sqrt{2}}$ & $0$ & $0$ & $0$ & $-1$
\\ \hline
 $8$ & $0$ & $\frac{1}{\sqrt{2}}$ & $\frac{1}{\sqrt{2}}$ & $0$ & $-\frac{1}{\sqrt{2}}$ & $\frac{1}{\sqrt{2}}$ & $-1$ & $0$ & $0$
\\ \hline
 $9$ & $\frac{1}{\sqrt{2}}$ & $0$ & $-\frac{1}{\sqrt{2}}$ & $-\frac{1}{\sqrt{2}}$ & $0$ & $-\frac{1}{\sqrt{2}}$ & $0$ & $1$ & $0$
\\ \hline
 $10$ & $0$ & $-\frac{1}{\sqrt{2}}$ & $\frac{1}{\sqrt{2}}$ & $0$ & $\frac{1}{\sqrt{2}}$ & $\frac{1}{\sqrt{2}}$ & $1$ & $0$ & $0$
\\ \hline
 $11$ & $\frac{1}{\sqrt{2}}$ & $0$ & $\frac{1}{\sqrt{2}}$ & $\frac{1}{\sqrt{2}}$ & $0$ & $-\frac{1}{\sqrt{2}}$ & $0$ & $-1$ & $0$
\\ \hline
 $12$ & $\frac{1}{\sqrt{2}}$ & $-\frac{1}{\sqrt{2}}$ & $0$ & $-\frac{1}{\sqrt{2}}$ & $-\frac{1}{\sqrt{2}}$ & $0$ & $0$ & $0$ & $1$   
\\ \hline
\end{tabular}
\caption{The basis vectors of one dimensional irreducible group
 representations of the space group $P4_132$ appearing in the 
 magnetic representation with $\vec{k}=(0,0,0)$.}
\label{tab:basis}
\end{table}

\begin{figure} 
\centering
 	\includegraphics[width=2.3in]{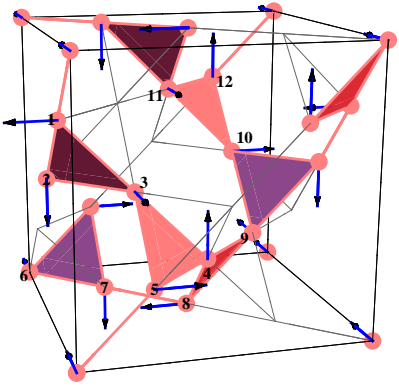}
	\caption{(Color online) The basis vector $\psi_3$ in Table.~\ref{tab:basis}.}
	\label{fig:hkorder2}
\end{figure}

The physical interpretation of these representations is as follows.  The
basis vector $\psi_1$ is nothing but the C$_2$ rotation axis at every
magnetic ion.  The basis vector $\psi_2$ gives the spin directions of
classical uncanted windmill state discussed above (see
Fig.~\ref{fig:hkorder1}).  The third basis vector $\psi_3$ may be
obtained as the axis which is
normal to both C$_2$ axis and the spin direction in $\psi_2$ (see
Fig.~\ref{fig:hkorder1}).  Note that these three basis vectors at each
site form an orthonormal basis for the spin coordinates.

Evidently $\Gamma_2$ is related to the DM
interaction, at least to the $D_2$ component. But what about $D_1$ and
$D_3$? Let us consider following situation. Starting from an ordered
ground state with $D_2<0$, we turn on an infinitesimal $D_1$ or $D_3$
component. The spin at site $i$ can be written as
\begin{equation}
{\bf S}_i = \sqrt{1-(a_i^1)^2-(a_i^3)^2} \hat{e}_i^2 + a_i^1 \hat{e}_i^1
+ a_i^3 \hat{e}_i^3  
\;,
\label{eq:spincorr}
\end{equation}
where $\hat{e}_1$, $\hat{e}_2$ and $\hat{e}_3$ are simply the three
orthogonal unit vectors given by basis $\psi_1$, $\psi_2$ and $\psi_3$,
and $a_i^1$ and $a_i^3$ are small corrections to the ordered ground
state due to the introduction of an infinitesimal $D_1$ (or $D_3$)
component.  We plug Eq.~\ref{eq:spincorr} into the Hamiltonian, and
expand to the $2nd$ order in $a_i^1$, $a_i^3$ and $D_1$ (or $D_3$).To
linear order $a_i^3$ vanishes. Thus the ground state spin configuration
with negative $D_2$ and an infinitesimal $D_1$ or $D_3$
component is related to $\psi_2$ and $\psi_3$. The irreducible
representation $\Gamma_2^{(1)}$ is relevant to the magnetic structure
when the DM interaction is present.

Now, we proceed by assuming that the ground state configuration $\psi$
of the more general case, when $D_1$, $D_2$ and $D_3$ are all present in
the system, is a linear superposition of basis vectors $\psi_2$ and
$\psi_3$:
\begin{equation}
{\bf S}_i = \cos{x} {\boldsymbol \psi}_{2,i} + \sin{x} {\boldsymbol \psi}_{3,i}
\;,
\label{eq:super}
\end{equation}
where ${\boldsymbol \psi}_{a,i}$ is the vector $\psi_a$ corresponding to
the $i^{\rm th}$ spin in Table~\ref{tab:basis}.  Evaluating
Eq.~\eqref{eq:hamiltonian} for spin configurations of this form gives
\begin{eqnarray}
{\mathcal H}/N
      & = & 2(-3 \sqrt{2} D_1 + 5 \sqrt{3} D_2-\sqrt{6} D_3 - 3 J) 
        \nonumber \\
      & & + 2 \sqrt{3} [( \sqrt{6} D_1 + D_2 + \sqrt{2} D_3 - \sqrt{3}J) \cos{(2x)} 
        \nonumber \\
      & & - (\sqrt{3} D_1 - 3 D_3)\sin{(2x)}]
\;,
\label{eq:angleeq}
\end{eqnarray}
where $N$ is the number of unit cells in the lattice, not the number of
spins (which is equal to $12N$). Minimizing the Hamiltonian with resepct to $x$, we
can find the canting angle $x$ is given by
\begin{eqnarray}
     \cos{(2x)}  & = & - 
    \frac{-\sqrt{3}J + \sqrt{6} D_1+D_2+\sqrt{2}D_3}{W}
         \\
  \sin{(2x)}  & = & \frac{\sqrt{3} D_1-3 D_3}{W} 
  \;,
\label{eq:canting}
\end{eqnarray}
where we have defined
\begin{equation}
W=\sqrt{(\sqrt{6}D_1+D_2+\sqrt{2}D_3-\sqrt{3}J)^2+3(D_1-\sqrt{3}D_3)^2}
\end{equation}
for convenience. 

\begin{figure}
 \centering
 	\includegraphics[width=2.5in]{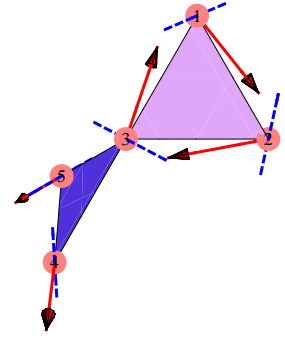}
	\caption{(Color online)The spin configuration of site $1,2,3,4,5$. 
The dashed blue lines are the canting axis of corresponding spin.}
	\label{fig:windmill}
\end{figure}

Fig.~\ref{fig:windmill} is an example of this canted state when $D_1 =
0.1J$, $D_2 = -0.04J$ and $D_3=0$.  We only plotted the spin
configuration of $\Delta123$ and $\Delta345$ in
Fig.~\ref{fig:dm}. Because these states are obtained by smoothly
introducing a $\psi_3$ component into the uncanted windmill states found
in Sec.~\ref{sec:order}, we will call this state a canted windmill
state.  The canted moment disappears not only when $D_1=D_3=0$, but also
for $D_1=\sqrt{3}D_3$, at which point it degenerates into the uncanted
windmill state.  Regardless, it is also interesting to note that because
$D_1,D_3 \ll J$, from Eq.~\eqref{eq:canting} the canted moment is much
smaller than the coplanar component, which indicates the dominance of
the coplanar spin configurations.  Similar features have been found in
other studies.

In the above treatment, we have assumed the ansatz in
Eq.~(\ref{eq:super}), which is not guaranteed to find the global minimum
energy state, and gives no guidance as to where this form of the ground
state breaks down. In next section, we will consider this question from
a different point of view.

\subsection{Mean field spherical model}
\label{sec:mft}

In this subsection we approach the general problem of finding classical
ground states of the Hamiltonian from a different point of view.  The
problem is difficult because in addition to minimizing ${\mathcal H}$,
which is quadratic in spins, we must also satisfy constraints that each
spin have fixed magnitude $|{\bf S}_i|=1$.  The large number (equal to
the number of spins) of these constraints makes what otherwise would be
a simple quadratic minimization problem difficult.  Here we replace
these many constraints by a single one,
\begin{equation}
  \label{eq:20}
  \sum_i |{\bf S}_i|^2 = 12N,
\end{equation}
where, as elsewhere in the text, we define $N$ as the number of unit
cells for convenience.  This is the ``spherical model'', and is exactly
soluble at both zero and non-zero temperature.  At zero temperature, the
spherical approximation {\sl must} give a lower bound to the true ground
state energy, since minimization is conducted with less constraints than
in the physical spin model.  Because of this observation, this approach
can indeed often be used to construct physical ground states.  This
``Luttinger-Tisza'' method.\cite{lyons:pr,luttinger:pr1,luttinger:pr2}
consists of finding a subset of ground states of the spherical model
which respect the spin normalization constraints of the physical
problem.  Any such states must be ground states of the full Hamiltonian.
Moreover, when such states exist, they exhaust the full set of physical
ground states.  However, it is not always possible to find any ground
states of the spherical model which satisfy the normalization
constraint.  If not, it simply means that ground state energy of the
physical problem is strictly larger than that of the spherical model,
and the Luttinger-Tisza method fails.  Generally, the Luttinger-Tisza
construction is less effective on lattices with a large number of sites
in their basis.  For the \hk\ lattice with a 12 site basis, our
expectations should not too high!  Nevertheless, in some range of phase
space, we will indeed find physical ground states from this approach.
More generally, at non-zero temperature, the spherical model may be a
useful approximation even when it fails to produce exact ground states
at zero temperature.

Minimizing the quadratic Hamilton in Eq.~(\ref{eq:hamiltonian}) with the
single global constraint in Eq.~(\ref{eq:20}) is a standard problem,
which is solved by finding the eigenvectors of the Hamiltonian matrix
(coefficients of the quadratic form of spin components) with minimum
eigenvalues.  By translational invariance, the eigenfunctions have the
Bloch form, i.e. are quasimomentum eigenstates.  Hence it is useful to
Fourier transform Eq.~\eqref{eq:hamiltonian}:
\begin{equation}
 {\mathcal H} = N \sum_{\bf k} \sum_{i,j} \sum_{\nu,\mu} L_{\nu \mu}^{ij}
               ({\bf k}) Q^{\nu \dagger}_i({\bf k}) Q^{\mu}_{j}({\bf k})
\;.
\label{eq:energy}
\end{equation}
Here
\begin{equation}
{\bf S}_i^{\nu}({\bf R}_n) = \sum_{\bf k}Q^{\nu}_i({\bf k})\exp{(i{\bf k} \cdot {\bf R}_n)}
\end{equation}
with $\nu$, $\mu$ index of spin vector components, ${\bf R}_n$ is the
position of unit cell, $i$ and $j$ are the sublattice index and $L_{\nu
\mu}^{ij}$ is the Fourier transformed Hamiltonian matrix in the Bloch
representation (which is $12\times 3=36$ dimensional because of the
multiple basis sites and spin components). We need to minimize
Eq.~\eqref{eq:energy} subject to the soft constraint
Eq.~\eqref{eq:spherical}, which can be expressed as
\begin{equation}
  \sum_{i=1}^{12} \sum_{\bf k}{\bf Q}_i^{\dagger}({\bf k})\cdot {\bf Q}_i({\bf k}) = 12
  \;.
  \label{eq:spherical}
\end{equation}

\begin{figure}
	\centering
	\subfigure{
	\label{fig:ph12}
		\includegraphics[width=3.0in]{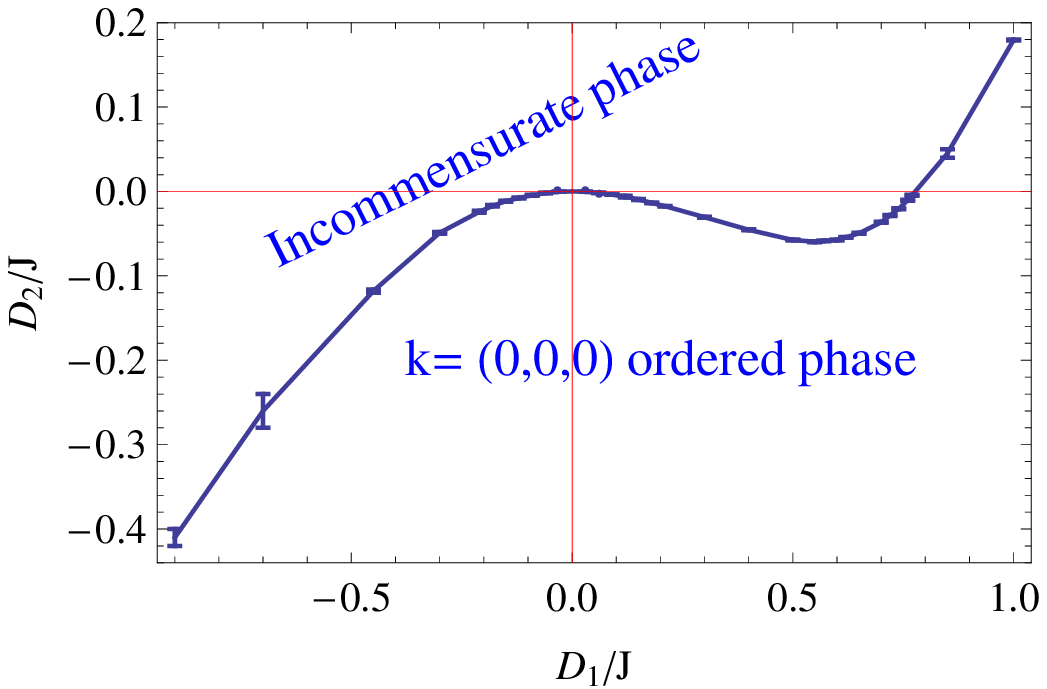}}
	\subfigure{
	\label{fig:ph23}
		\includegraphics[width=3.0in]{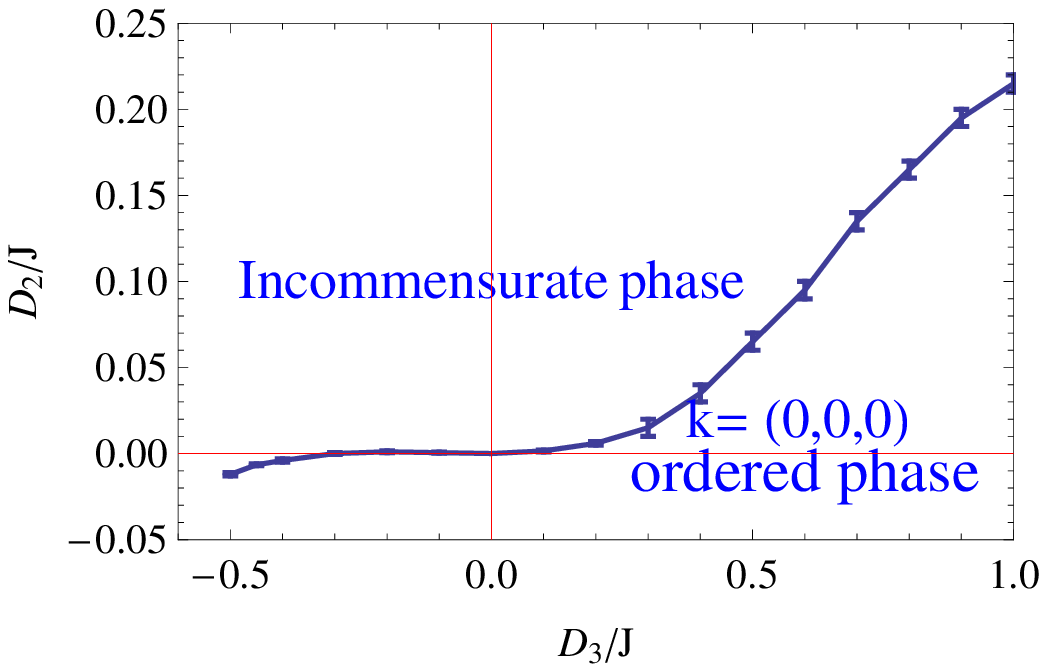}}
	\subfigure{
	\label{fig:ph31}
		\includegraphics[width=3.0in]{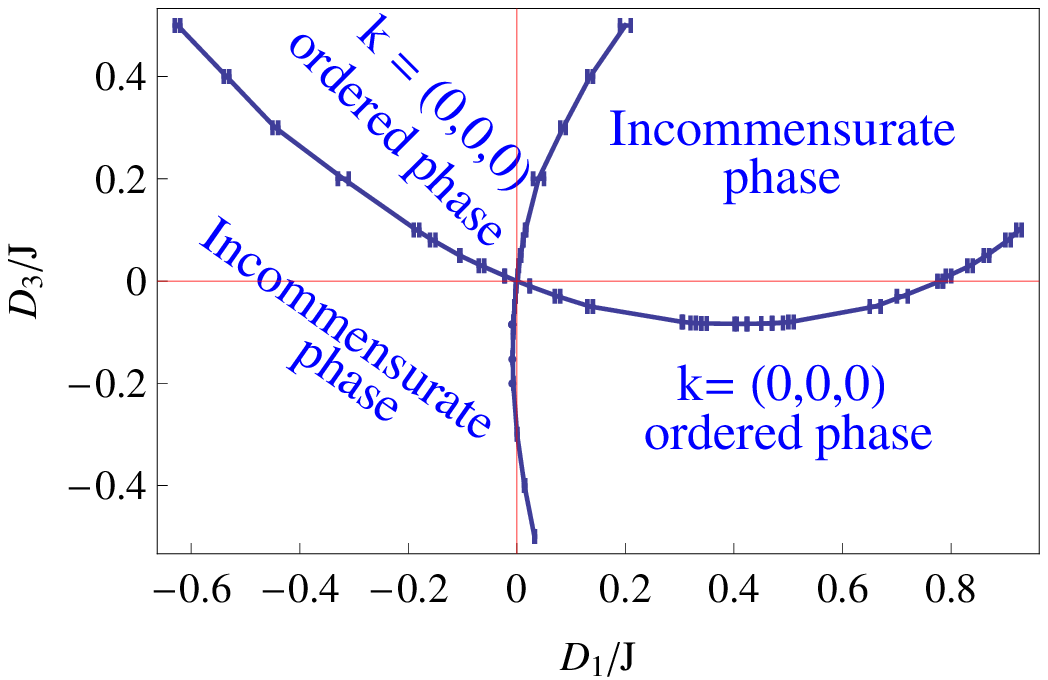}}
  	\caption{(Color online) phase diagram in $D_1$-$D_2$, $D_2$-$D_3$ and $D_1$-$D_3$ 
  	parameter spaces. The uninvolved DM vector component in each figure is set to be $0$. 
  	The red reference lines (axes) are not phase boundary.}
	\label{fig:pd}
\end{figure}

Minimization is equivalent to find the minimum eigenvalues (and
corresponding eigenvectors) of $L_{\nu \mu}^{ij}({\bf k})$.  We did this
numerically for every ${\bf k}$, and found the global minimum for every
$(D_1,D_2)$, $(D_2,D_3)$ and $(D_1,D_3)$ pairs. With this approach,
phase diagrams in $D_1$-$D_2$, $D_2$-$D_3$, $D_1$-$D_3$ parameter spaces
have been plotted in Fig.~\ref{fig:pd}.  In a wide region of the phase
diagrams, the minimum eigenvalue is realized for ${\bf k}=(0,0,0)$.  In
this case, the corresponding eigenfunction {\sl can} be chosen to
satisfy the normalization constraint on every site, and so an exact
ground state is found.  This ordered state is in fact precisely the
canted/uncanted windmill states we proposed in previous
section.  Thus in the regions for which ${\bf k}=(0,0,0)$ is indicated
in the figures, this analysis proves that these windmill states are the
exact global ground states. In a considerable large regions of the 
parameter spaces, we get canted/uncanted ``windmill'' states.

In other broad regions of the phase diagram, the spherical model
predicts ordered states with incommensurate wavevectors, i.e. in which
${\bf k}$ has irrational projection onto reciprocal lattice vectors.
This is indicated simply as ``incommensurate phase'' in the figures.  In
most cases we have studied, the incommensurate wavevectors are located
around $0.85(\pi,\pi,\pi)$ and its eight equivalent momenta $0.85(\pm
\pi, \pm \pi, \pm \pi)$.  However, in this region of the phase diagram,
we are unable to construct a linear combination of eigenfunctions which
satisfies the local constraint on the spin magnitudes.  Thus the
incommensurate ground state of the spherical model does not immediately
imply a corresponding ground state of the physical model.  It is
possible that the region of phase space occupied by the windmill states
is actually expanded beyond what is shown here by this effect.  Most
likely, ground states with large unit cells or incommensurate order do
exist in the physical model, but are more complex than those of the
spherical approximation, and with somewhat higher energy.  Even in the
spherical model, we see that in the incommensurate region, while the DM
interaction removes much of the frustration-induced degeneracy, the
enlargement of the unit cell implies a larger residual ground state
degeneracy, and hence less effective removal of frustration than in the
${\bf k}=(0,0,0)$ regions.

\section{Quantum effects}
\label{sec:spinwave}

\subsection{Numerically constructed Bogoliubov transformation}
\label{sec:nb}
   
In previous sections, the spins were treated classically and classical 
ground states were obtained. In this section, we discuss 
the quantum effect in the formalism of linear spin wave theory. In 
certain regions of $D_1$-$D_2$, $D_2$-$D_3$ and $D_1$-$D_3$ parameter 
space, we have an ordered ground state. We will use the Holstein-Primakoff 
Boson approach to explore the quantum effects.\cite{hp:pr} Now express the 
spin operator as follows,
\begin{eqnarray}
{\bf S}_i({\bf R}_n) \cdot \hat{S}_i^{ord} & \simeq & S - a_i^{\dagger}({\bf R}_n)a_i({\bf R}_n)  \\
{\bf S}_i({\bf R}_n) \cdot {\bf e}_i^1 & \simeq & \frac{\sqrt{2S}}{2I}(a_i({\bf R}_n) -a_i^{\dagger}({\bf R}_n)) \\
{\bf S}_i({\bf R}_n) \cdot (\hat{S}_i^{ord} \times {\bf e}_i^1) & \simeq & \frac{\sqrt{2S}}{2}(a_i({\bf R}_n)+a_i^{\dagger}({\bf R}_n)) 
\;,
\label{eq:boson}
\end{eqnarray}
where $\hat{S}_i^{ord}$ is the unit vector along the spin order, 
${\bf e}_i^1$ is the $C_2$ rotational axis at site $i$ introduced in Sec.~\ref{sec:representation}, $a_i^{\dagger}$ and $a_i$ are the 
creation and annihilation operators of Holstein-Primakoff bosons 
at $i$th sublattice of unit cell at position ${\bf R}_n$, and we 
only keep the lowest order of $a_i^{\dagger}$ and $a_i$. Under 
this transformation, the Hamiltonian can be written as 
\begin{eqnarray}
{\mathcal H} 
    &=&  - \frac{S}{2}\sum_{i,j,{\bf k}} A_{ij}({\bf k})a_i^{\dagger}
         ({\bf k})a_j({\bf k}) +B_{ij}({\bf k}) a_i^{\dagger}({\bf k})a_j^{\dagger}
         (-{\bf k})  
       \nonumber \\
    &+&  H.c. \;,
\label{eq:swham}
\end{eqnarray}
where we have dropped the constant term and high order terms. 
Here, $A_{ij}({\bf k})$ and $B_{ij}({\bf k})$ are the coefficient 
matrix we end up with after doing Fourier's transform on the 
creation and annihilation operators. The Fourier's transform we used is
\begin{eqnarray}
a_i^{\dagger}({\bf R}_n) 
  & = & \frac{1}{\sqrt{N}} \sum_{\bf k} a_i^{\dagger}\exp{ ( i{\bf k} \cdot {\bf R}_n )}
  \\
a_i({\bf R}_n) 
  & = & \frac{1}{\sqrt{N}} \sum_{\bf k} a_i \exp{( -i{\bf k} \cdot {\bf R}_n )}
  \;.
\end{eqnarray}

Since there are $12$ sublattices, using the analytical Bogoliubov 
transformation is hopeless to diagonalize the Hamiltonian. Here, 
we will use a numerically constructed Bogoliubov transformation 
(NCBT) introduced and discussed in detail by 
Ref.~\onlinecite{Maestro:jpcm,Maestro:03} to diagonalize Eq.~\eqref{eq:swham}, 
find the spin 
wave energy gap and calculate the quantum corrections to the classical 
order. Write Eq.~\eqref{eq:swham} as 
\begin{equation}
{\mathcal H} = \sum_{\bf k} {\bf X}^{\dagger}({\bf k}) {\bf H}({\bf k}) {\bf X}({\bf k})
\;,
\label{eq:matform}
\end{equation}
where 
\begin{eqnarray}
{\bf X}({\bf k}) 
    &=& (a_1({\bf k}) \ldots a_{12}({\bf k}), a_1^{\dagger}(-{\bf k}) \ldots
     a_{12}^{\dagger}(-{\bf k}))^T 
      \\
{\bf H}({\bf k}) 
    &=& -\frac{S}{2}\left( \begin{array}{cc} A({\bf k}) & B({\bf k})
      \\ B^{\ast}(-{\bf k}) & A^{\ast}(-{\bf k}) \end{array} \right) \;,
\label{eq:mat2}
\end{eqnarray}
and the hermiticity of ${\mathcal H}$ requires that 
\begin{eqnarray}
A_{ij}({\bf k}) &=& A^{\ast}_{ij}({\bf k}) \\
B_{ij}({\bf k}) &=& B_{ji}({\bf k}) \;.
\label{eq:condition}
\end{eqnarray}
We now introduce the canonical transformation 
\begin{equation} 
{\bf X}({\bf k}) = {\bf Q}({\bf k}){\bf Y}({\bf k})
  \;,
\label{eq:canon}
\end{equation}
where, ${\bf Y}({\bf k})$ is given by
\begin{equation}
  {\bf Y}({\bf k})  =  
  (b_1({\bf k}) \ldots b_{12}({\bf k}), b_1^{\dagger}(-{\bf k}) \ldots b_{12}^{\dagger}(-{\bf k}))^T  
  \;,
\end{equation}
and satisfies
\begin{equation}
  [ b_i({\bf k}), b_j^{\dagger}({\bf k}^{\prime}) ]  = \delta_{ij} \delta_{{\bf k},{\bf k}^{\prime}}
\;.
\label{eq:yop}
\end{equation}
The transformation ${\bf Q}$ is required to to diagonalize the
Hamiltonian as
\begin{equation} 
{\bf Q}^{\dagger}({\bf k}){\bf H}({\bf k}){\bf
    Q}({\bf k}) = {\bf \Lambda}({\bf k}) 
    \;, 
\label{eq:diag}
\end{equation}
where ${\bf \Lambda}({\bf k})$ is the diagonal eigenvalue matrix whose
diagonal matrix elements are given by $(\epsilon_1({\bf
  k}),\cdots,\epsilon_{12}({\bf k}),\epsilon_1({-\bf
  k}),\cdots,\epsilon_{12}({-\bf k}))$.  Using this transformation, the
quantum correction to the classical spin polarization can be written
as
\begin{eqnarray}
  dS 
  &=& \frac{1}{12 N}\sum_{n,i} \langle a_i^{\dagger}({\bf R}_n) a_i({\bf R}_n) \rangle \nonumber \\
  &=& \frac{1}{12 N}\sum_{{\bf k}, i} \langle a_i^{\dagger}({\bf k}) a_i({\bf k})  \rangle \nonumber \\
  &=& \frac{1}{24 N} \sum_{\bf k} \langle  {\bf X}^{\dagger} {\bf X} \rangle - \frac{1}{2} \;,
\label{eq:qc1}
\end{eqnarray}
At zero temperature, further making use of Eq.~\eqref{eq:canon},
Eq.\eqref{eq:qc1} can be expressed as
\begin{equation}
  dS = \frac{1}{2} \{ \frac{1}{12N} \sum_{\bf k} \sum_{i=1}^{12} [{\bf Q}^{\dagger}{\bf Q}]_{ii} -1  \} 
  \;.
\label{eq:qc2}
\end{equation}
If we find the canonical transformation ${\bf Q}({\bf k})$, the energy
spectrum can also be obtained. With the energy spectrum, we can find the
spin wave energy gap, $\Delta$.  Some care must be taken as the
numerical construction of the Bogoliubov transformation is effective
only when there is an energy gap.

\subsection{Quantum corrections and spin wave gaps}

\begin{figure}
	\centering
	\subfigure{
	\label{fig:d10d2qc}
		\includegraphics[width=3.0in]{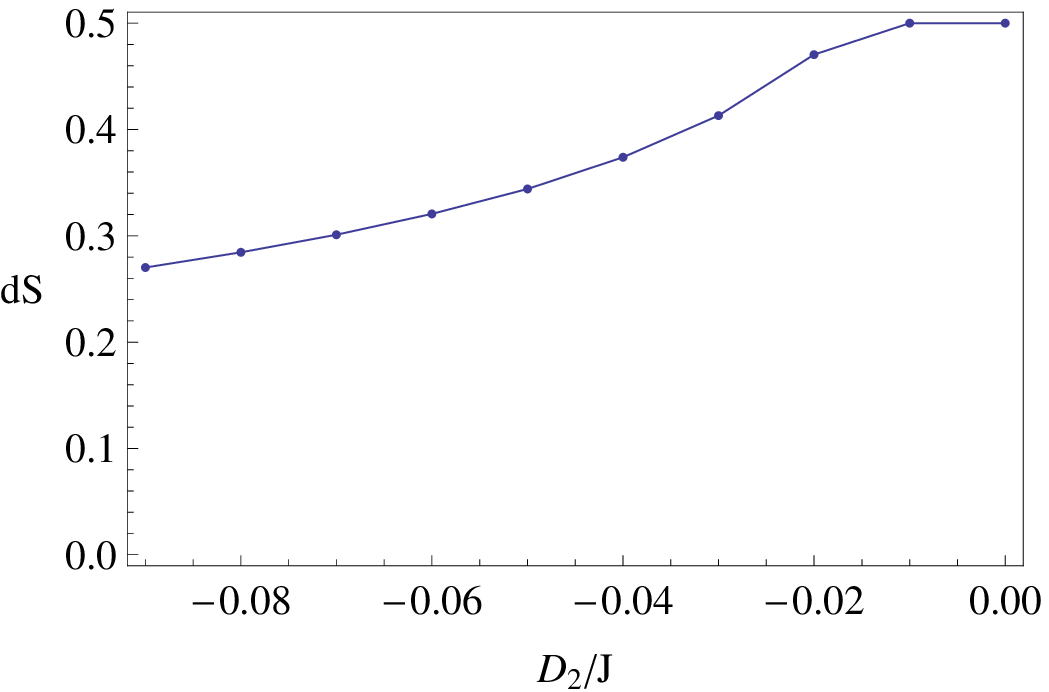}}
	\subfigure{
		\includegraphics[width=3.0in]{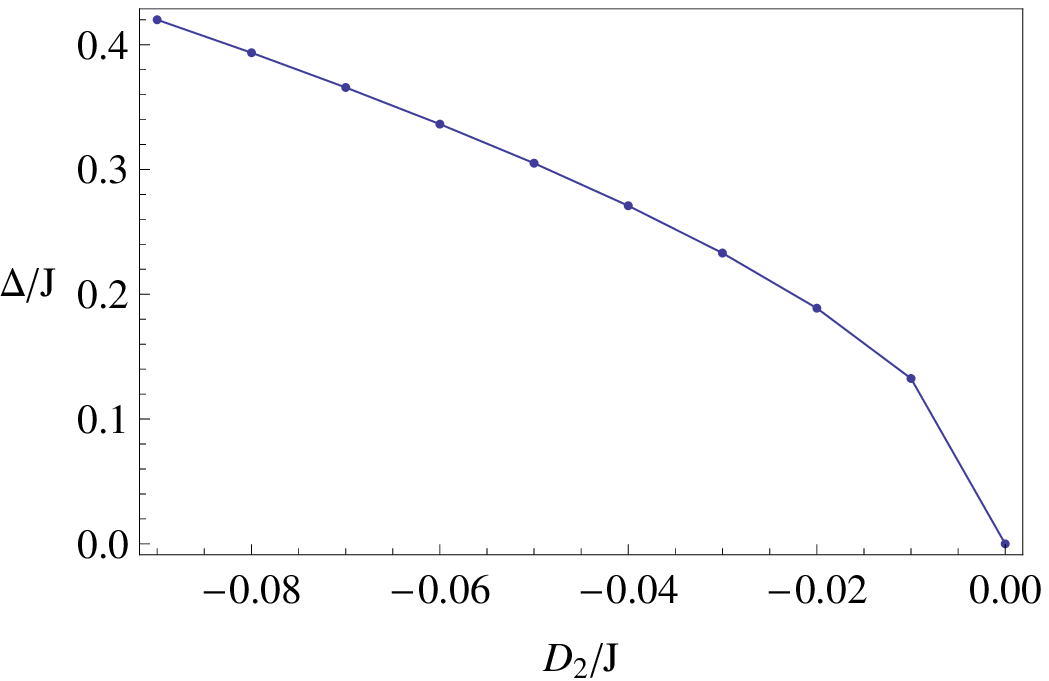}}
  	\caption{(Color online) The dependence of quantum corrections and spin-wave gaps on the DM vector components. In the two figures, we set $D_1=D_3=0$ and vary $D_2$. $60\times60\times60$ momentum points have been used to generated the data. No change has been found in quantum corrections and gaps within computer resolution compared with $50\times50\times50$ momentum points.}
	\label{fig:qcgap1}
\end{figure}

\begin{figure}   
   	\subfigure{
		\includegraphics[width=3.0in]{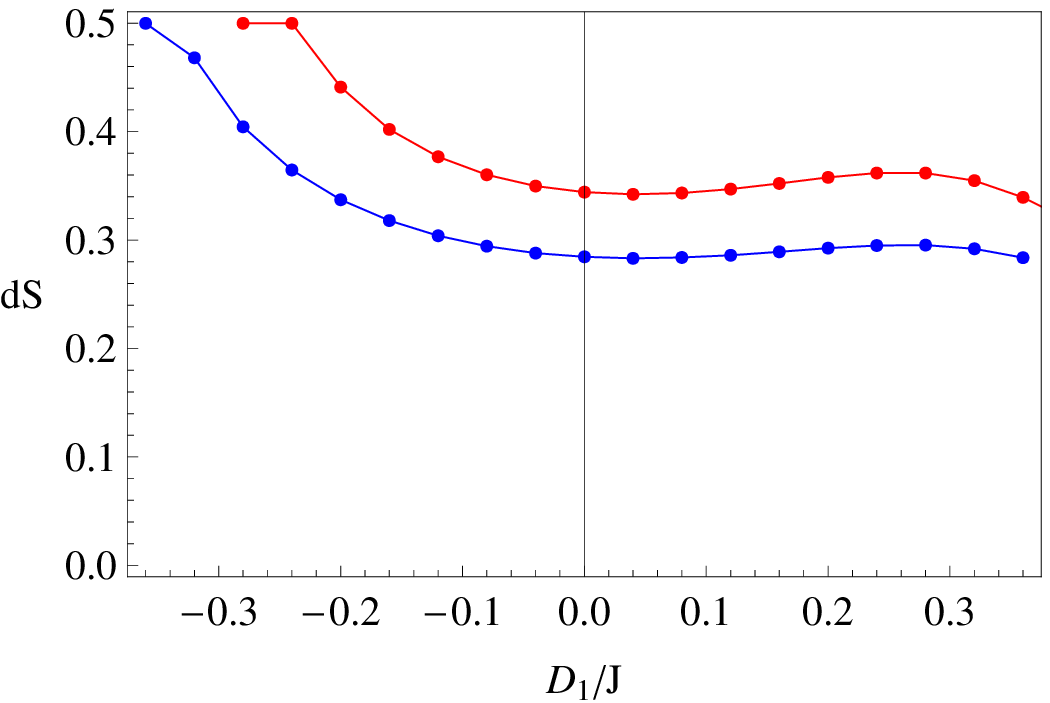}}
	\subfigure{
		\includegraphics[width=3.0in]{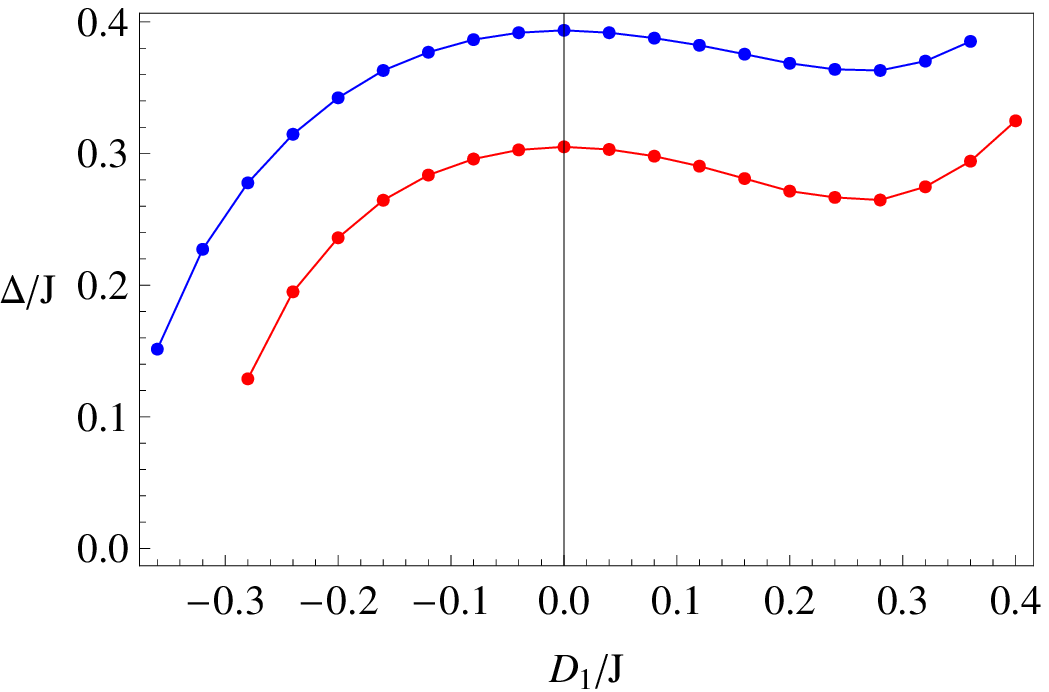}}  
	\caption{(Color online) The dependence of quantum corrections and spin-wave gaps on the DM vector components. In the two figures, we set $D_3=0$ and vary $D_1$ with two fixed $D_2$ values ($D_2=-0.08J$(in blue) and $D_2=-0.05J$(in red)). $16\times16\times16$ momentum points have been used to generated the data. No change has been found in quantum corrections and gaps within $1\%$ compared with $10\times10\times10$ momentum points (Same for Fig.~\ref{fig:qcgap3} and Fig.~\ref{fig:qcgap4}).} 
	\label{fig:qcgap2}
\end{figure}

\begin{figure}
	\centering
	\subfigure{
		\includegraphics[width=3.0in]{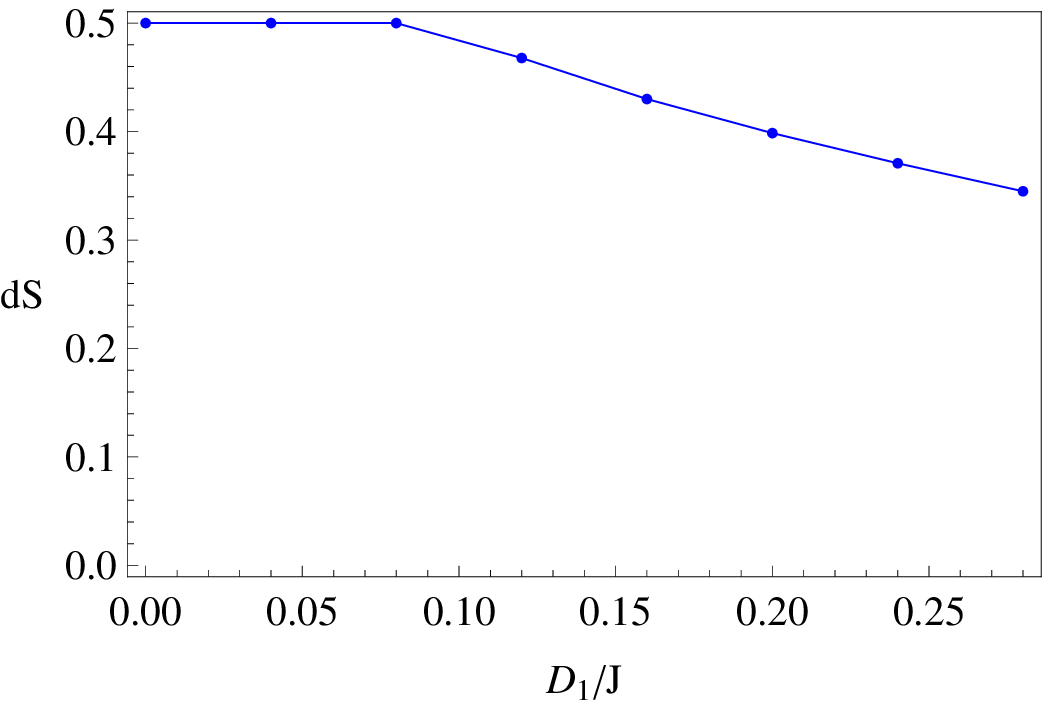}}
	\subfigure{
		\includegraphics[width=3.0in]{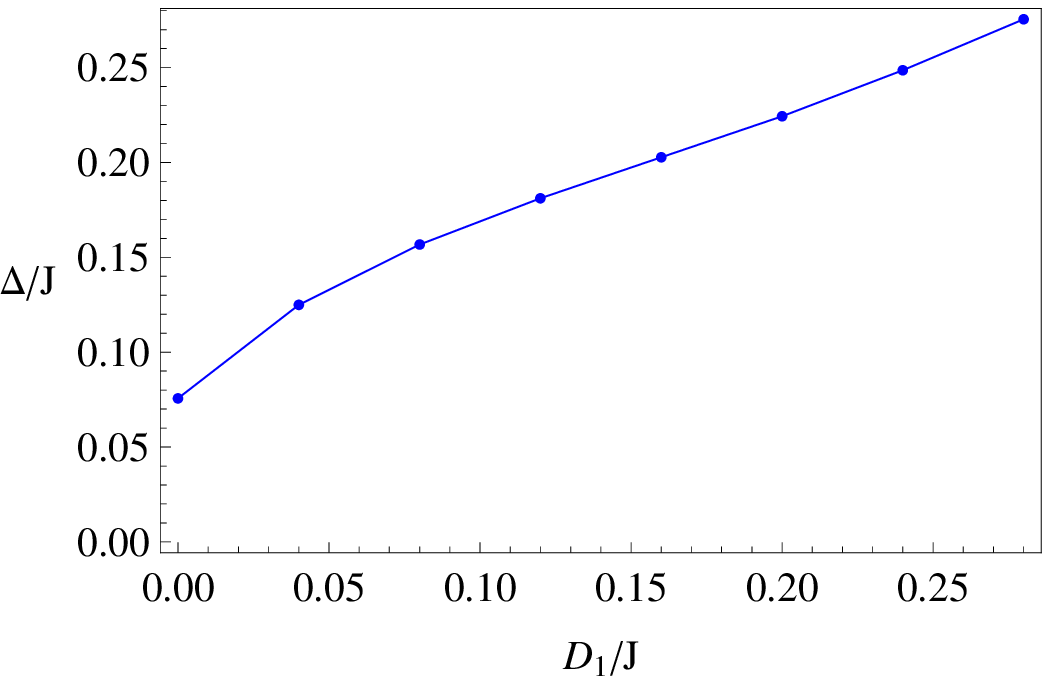}}
  	\caption{(Color online) The dependence of quantum corrections and spin-wave gaps on the DM vector components. In the two figures, we set $D_2=0,\ D_3=-0.1$ and vary $D_1$. We also did the same thing with $D_2=0,\ D_3=0.1$ and varying $D_1$, and the quantum corrections always break the classical order completely.}
	\label{fig:qcgap3}
\end{figure}

\begin{figure}
   	\subfigure{
		\includegraphics[width=3.0in]{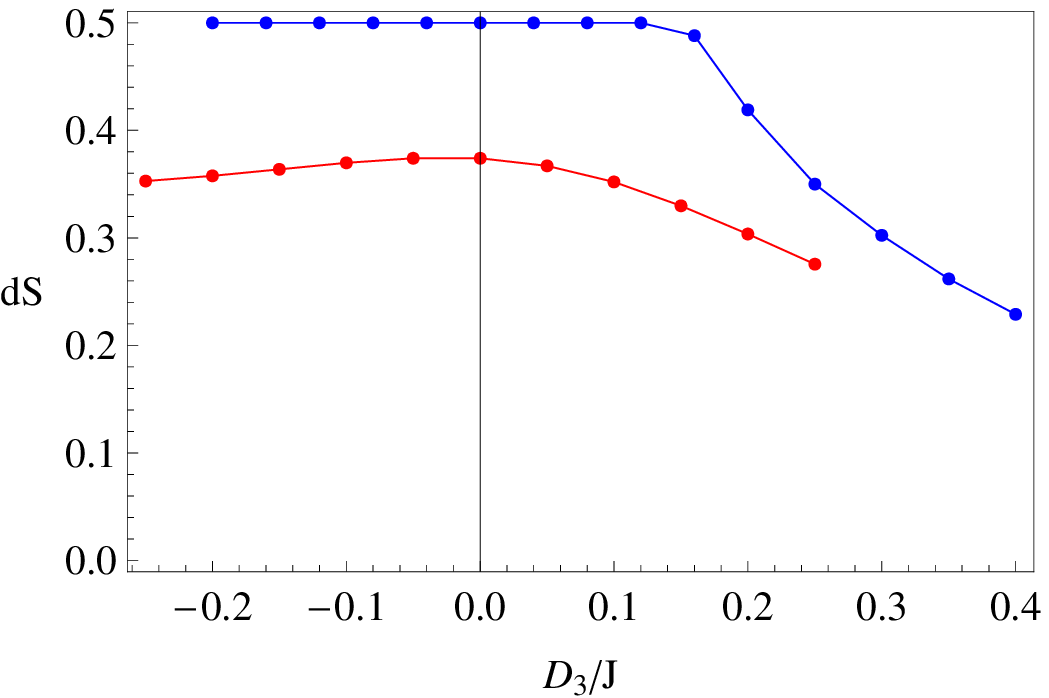}}
	\subfigure{
		\includegraphics[width=3.0in]{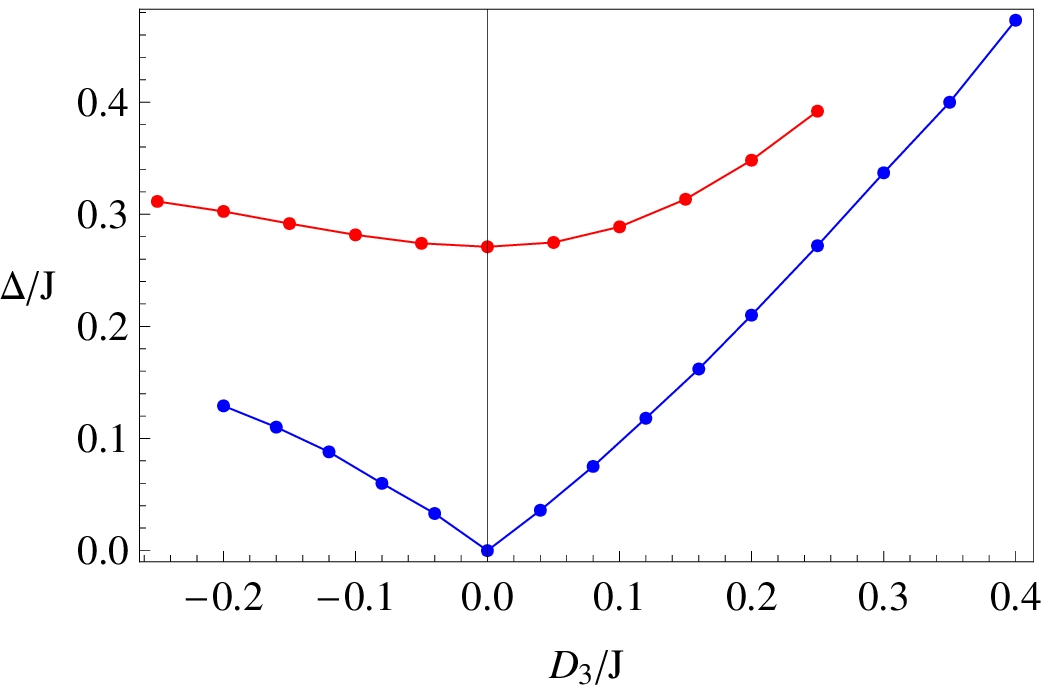}}	
	\caption{(Color online) The dependence of quantum corrections and spin-wave gaps on the DM vector components. In the two figures, we set $D_1=0$ and and vary $D_3$ with two $D_2$ values ($D_2=0$(in blue) and $D_2=-0.04J$(in red)). We also did the same thing with $D_2=0,\ D_3=0.1$ and varying $D_1$, and the quantum corrections always break the classical order completely.}
	\label{fig:qcgap4}
\end{figure}

By the method described in last section, we carried out the numerical
procedure described in Ref.~\onlinecite{Maestro:jpcm,Maestro:03}.
Taking spin $S=\frac{1}{2}$, we numerically construct the Bogoliubov
transformation for every ${\bf k}$, and find its contribution to zero
temperature quantum correction, $dS$, and energy levels at every ${\bf
  k}$ to extract the spin wave gaps. The numerical results are plotted
in Fig.~\ref{fig:qcgap1}, Fig.~\ref{fig:qcgap2}, Fig.~\ref{fig:qcgap3}
and Fig.~\ref{fig:qcgap4}.  Corrections $dS$ larger than $1/2$ have been
truncated to $1/2$.  In these figures, two components of the DM
vector are kept constant while the third is varied. The ordered regions
(the third varying DM vector component) of these figures can be found
the Fig.~\ref{fig:pd}.
   
In these figures, spin wave gap is non-vanishing, so our calculation
is valid. It is easy to find the quantum behavior also resembles the
classical one: the different DM vector components have different
effects in quantum corrections, which is similar to the effect of DM
vector components in favoring canted ``windmill'' state in
Sec.~\ref{sec:representation}. In the DM magnitude studied in these
figures, the quantum corrections are pretty large. Even in the case
when $D_2=-0.09J$ and $D_1=D_3=0$, the quantum correction is about $50
\%$.  

As a general rule, one observes that the quantum corrections decrease
steadily as one goes deeper into the ${\bf k}=0$ classically ordered
region.  If we crudely suppose that $dS>1/2$ is indicative of the
destruction of order by quantum fluctuations, we may expect broad
regions of quantum spin liquid states occurring in and near the
incommensurate regions of the classical phase diagram.  This range of
DM vectors then may be possible candidates for application to \nio.  

\subsection{Comparison with exact diagonalization}

\begin{figure}
 \centering
 	\includegraphics[width=3.0in]{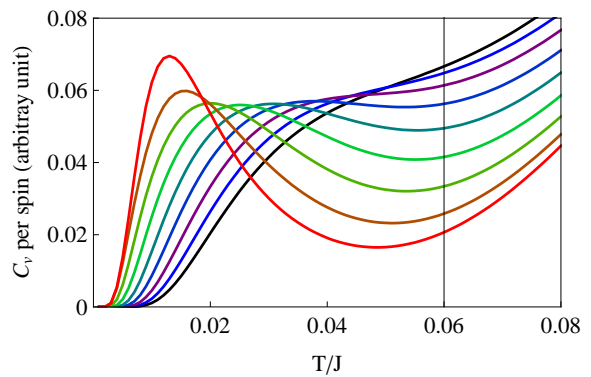}
	\caption{(Color online) The specific heat of six triangle 
	with negative $D_2$ component DM interaction. Along the thin vertical
	line, from top to bottom $D_2$ value of each curve increases from $-0.09J$ to $-0.01J$
	with a step $0.01J$.}
	\label{fig:cv6d2}
\end{figure}

In order to partially confirm our results in last section, we performed
numerical exact diagonalization for $S=1/2$
spins.\cite{alet:jpsjs,albuq:jmmm} We took six triangles with thirteen
sites and used a Heisenberg model plus DM interaction with only $D_2\neq
0$.  We plot the resulting specific heat in Fig.~\ref{fig:cv6d2}. The
gap in each case can be inferred from the plot by the temperature below
which the specific heat becomes negligible.  As we found in
previous section, the more negative $D_2$ is, the greater the gap will
be. At low temperatures in Fig.~\ref{fig:cv6d2}, the sequence of the
curves agrees with what they should behave according to spin wave gaps
in Fig.~\ref{fig:qcgap1}.

\begin{figure}
 \centering
 	\includegraphics[width=3.0in]{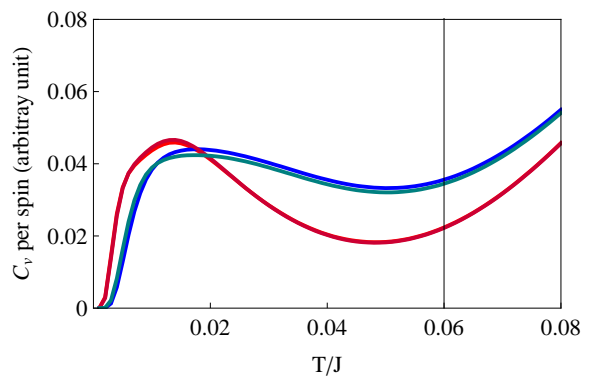}
	\caption{(Color online) The specific heat of six triangle 
	with $D_3$ component DM interaction. Along the thin vertical
	line, the upper curves have $|D_3| = 0.25J $, the down curves
	have $|D_3|= 0.10J$.}
	\label{fig:cv6d3}
\end{figure}

Similarly, we also look at the case when only $D_3$ component is 
present by taking $D_3 = \pm 0.10J$, $\pm 0.25J$. According to 
Fig.~\ref{fig:qcgap4}, the spin wave gaps of $D_3=\pm 0.10J$ 
are close to each other, and the spin wave gaps of $D_3 = \pm 0.25J$ 
are also close to each other, but much larger than the previous cases. 
In Fig.~\ref{fig:cv6d3}, we see that both curves of $D_3 = \pm 0.10J$ 
and $D_3 = \pm 0.25J$ nearly overlap at low temperatures, and their 
sequence agrees with the magnitudes of the spin-wave gaps. 

\section{Discussion}
\label{sec:discussion}

In this paper we have studied the effect of spin-orbit interactions in
the \hk\, lattice of \nio.  A crucial physical parameter is the strength
of atomic spin-orbit coupling relative to non-cubic crystal field
splittings.  In the strong spin-orbit limit, Ir-O-Ir superexchange
produces highly anisotropic effective spin interactions, with 2/3
antiferromagnetic and 1/3 {\sl ferromagnetic} couplings between spin
components.  This Hamiltonian turns out to be {\sl unfrustrated}, and
has a small set of classical ground states.  We speculated this even the
$S=1/2$ quantum model is likely ordered with a substantial critical
temperature, inconsistent with experiments on \nio.

By contrast, strong spin-orbit interactions and direct Ir-Ir exchange,
or weak spin orbit interactions, all induce an Heisenberg-like
Hamiltonian with a small correction of the form of a
Dzyaloshinskii-Moriya (DM) term.  The orientation of the DM vector,
which is not determined by symmetry or our microscopic considerations,
determines the extent to which the frustration of the Heisenberg model
is relieved.  In one region of phase space, frustration is fully
relieved, and the DM interaction stabilizes a magnetically ordered
``windmill'' state, with nearly but generically not quite coplanar
moments.  Quantum fluctuations, which we assessed by spin wave theory,
are sufficiently suppressed that we may expect this order to persist
even for spin-$1/2$ spins (as in \nio) in part of this region. In the
remainder of phase space, the frustration is not fully removed, and the
classical ground states break the lattice periodicity and may be
incommensurate.  We argued that in this regime, the classical ordering
is fragile and may be destroyed by quantum fluctuations for $S=1/2$
spins.

\subsection{Zero temperature susceptibility in quantum spin liquids}
\label{sec:zero-temp-susc}

Part of the motivation of the present study was the observation in \nio\
that the susceptibility $\chi$ tends to a constant at low temperature,
despite the approximately quadratic decrease of specific heat.  We
argued that this combination, which implies a diverging Wilson ratio as
$T\rightarrow 0$, is likely indication of spin-orbit interactions.
Indeed, on general grounds, a constant zero temperature susceptibility
is expected when SU(2) spin-rotation symmetry is broken.  The situation
of weak DM interaction is quite common in frustrated magnets, and may
allow this behavior quite broadly. Therefore it is interesting to
consider more generally how this occurs in the presence of weak DM
coupling.  We have not so far addressed the {\sl magnitude} of this zero
temperature susceptibility.

Presuming the DM interaction to be relatively weak, the magnitude of
$\chi$ should be understood in terms of the correlations the spins
would have in the underlying system without DM.  Various SU(2)
invariant phases lead to rather different behaviors.  Generally
speaking, one expects the most suppressed $\chi$ for systems with the
least low-energy spin fluctuations in the absence of DM.  Probably the
most extreme example is a Valence Bond Solid (VBS) or dimer state, in
which the eigenstates can be approximated by those of a single
partition of the sites into pairs of spins which are coupled to each
other only within the pairs.  Such a VBS phase has a gap of order $J$
to all excitations, including the elementary triplets. A simple
calculation by second order perturbation theory of the susceptibility
shows that it is indeed non-zero, and of order
\begin{equation}
  \label{eq:dvbs}
  \chi_{VBS}(T=0) \sim \frac{D^2}{J^3}
  \;.
\end{equation}
One may also estimate the magnitude of $\chi$ for various
phenomenological gapless spin liquid ground states perturbed by DM.
The general arguments follow scaling theory. We presume the gapless
spin liquid is a critical phase in the renormalization group sense,
described by a scale invariant field theory. Introduction of DM
interactions breaks SU(2) symmetry, and allows operators ${\mathcal
  O}_\alpha$ breaking SU(2) to be added to the effective
action/Hamiltonian. Generically, these appear with coefficients
proportional to $D$. In the simplest situation, there is a single
such operator ${\mathcal O}_{\Delta}$ with the smallest scaling
dimension $\Delta$. In most cases of interest, we expect
$\Delta<d+z$, where $d$ is the spatial dimension and $z$ is the
dynamical critical exponent ($z=1$ is common). In this case, the
presence of this operator in the Hamiltonian constitutes a {\sl
  relevant} perturbation. Then, if the susceptibility at $D=0$
behaves as $\chi \sim T^{\beta}$, we expect
\begin{equation}
   \chi_{\mu\nu}(D,T) \sim T^{\beta} f_{\mu\nu}(D/T^{\frac{d+z-\Delta}{z}})
   \;,
\label{eq:chiscale}
\end{equation}
where $\mu,\nu$ are spin components $x,y,z$. The operator ${\mathcal
O}_\Delta$ is expected to break SU(2) down to some subgroup.  This
may contain either one or zero residual U(1) spin rotation axes.  The
susceptibility normal to this axis, if it exists, is expected to be
constant at low temperature.  If no such axis exists, then the
susceptibility will be constant in all directions.  
In either case, we must have
\begin{equation}
  \label{eq:fscale}
  f_{\mu\nu}(X) \sim A_{\mu\nu} X^{\frac{\beta z}{d+z-\Delta}}, \qquad
  \mbox{for }|X|\gg 1.
\end{equation}
Here $A_{\mu\nu}$ is a symmetric tensor with either 2 or 3 non-zero
eigenvalues, in the cases with one or zero residual U(1) symmetries,
respectively.  One thereby obtains 
\begin{equation}
  \label{eq:chizero}
  \chi_{\mu\nu}(T=0) \sim |D|^{\frac{\beta z}{d+z-\Delta}} A_{\mu\nu}.
\end{equation}

As an example, consider the 2d ``Dirac'' spin liquid with point nodes
on the kagome lattice studied by Hermele {\sl et al} \cite{hermele:08}.
There, the dominant
operator indeed preserves a single residual U(1) symmetry. Its scaling
dimension is estimated as $\Delta \approx 2 - 32/(8\pi^2) \approx 1.6$
(based on a calculation for a generalized model with large number,
$N_f$, of flavors of fermions, evaluated for the physical case
$N_f=4$).  Taking $d=2,z=1,\beta=1$ as appropriate for this case, we
find, restoring units
\begin{equation}
  \label{eq:susc}
  \chi_\perp(T=0) \sim \frac{\mu_B^2}{J} \left|\frac{D}{J}\right|^{0.7}.
\end{equation}
Here $\chi_\perp$ is the susceptibility in the x-y plane perpendicular
to the conserved U(1) spin axis.  We see that the dependence on $D$ is
{\sl sub-linear}, making for a very large susceptibility even for
rather small $D/J$.  

It is noteworthy that the scaling prediction above should obtain
regardless of the other properties of the system in the presence of DM
interaction.  The relevance of ${\mathcal O}_\Delta$ at the spin
liquid fixed point indeed implies that it drives the system into a
different phase, which may not be a spin liquid at all.  This is
believed to be the case for the above Dirac spin liquid, for which the
resulting state is expected to be magnetically ordered.\cite{hermele:08}

\subsection{Other possibilities}
\label{sec:other-possibilities}

One may wonder whether the weak and strong spin orbit limits are the
only possibilities for \nio, and whether they might be distinguished
more directly.  Probably the principal difference in the two cases is
the {\sl sign} of the $g$-factor.  In the weak spin orbit limit, one has
approximately ${\bf M} \approx -2\mu_B {\bf S}$, while in the strong
case, we found ${\bf M} \approx +2\mu_B {\bf S}$.  While these lead to
identical Curie laws, they are physically distinct (note that one cannot
reverse the sign of ${\bf S}$ and maintain its canonical commutation
relations).  It should be measurable in other experiments such as
nuclear magnetic resonance.  Microscopic reasoning gives no reason why
the Ir$^{4+}$ spins might not be in an intermediate situation between
the two extreme limits.  However, in this case one would expect a
$g$-factor in between these two values, i.e. with substantially reduced
magnitude.  A large deviation would seem to be inconsistent with the
measured spin susceptibility.

This tends to support the notion that \nio\ is either in the strong or
weak spin orbit limit, and not in between.  Given the incompatibility of
the strongly anisotropic Ir-O-Ir superexchange Hamiltonian in the strong
spin-orbit case with experiment, we are led to believe the weakly
anisotropic Hamiltonian with DM interactions is most appropriate (we note that ``weak anisotropy'' still allows for $|D|/J
\sim 0.1$ which would have strong effects on the low energy physics).  This,
however, still leaves open the issue of weak versus strong spin-orbit
interactions. Though susceptibility experiments do not distinguish the
two cases, they are physically distinct, and could be discriminated by
magnetic resonance methods, for instance.  So far as we are aware, all prior
measurements of Ir$^{4+}$ ions capable of this distinction have been
interpreted in terms of the strong spin-orbit scenario (see for example,
Ref.\onlinecite{Raizman:prb}).  This fundamental physical question in
\nio\ warrants further investigation.

Could there be another scenario?  We cannot rule out the possibility
that other interactions might play a role.  Perhaps further neighbor
exchange or spin-lattice coupling might be significant.  These are
important subjects for future theoretical studies.

\begin{acknowledgments}

We are grateful to M.J.P. Gingras, Y.B. Kim, A.
Laeuchli, A. Schnyder, R.R.P. Singh, E.M. Stoudenmire for discussion.
We also wish to acknowledge fruitfull discussions with S. Trebst, A.E.
Feiguin and A.G. Del Maestro regarding some of the numerical work
presented in this paper.  This work was supported by a David and
Lucile Packard Foundation Fellowship and the NSF through DMR04-57440.

\end{acknowledgments}

\end{document}